\documentclass[namedreferences]{solarphysics}
\usepackage[optionalrh]{spr-sola-addons} 
\usepackage{graphicx}        
\usepackage{color}           
\usepackage{url}             

\newcommand{\aap}{    {\it Astron. Astrophys.}}

\newcommand{\aapr}{   {\it Astron. Astrophys. Rev.}}

\newcommand{\aj}{     {\it Astron. J.}} 
\newcommand{\apj}{    {\it Astrophys. J.}}
\newcommand{\apjl}{   {\it Astrophys. J. Lett.}}

\newcommand{\jgr}{    {\it J. Geophys. Res.}}

\newcommand{\pasj}{   {\it Pub. Astron. Soc. Japan}}

\newcommand{\solphys}{{\it Solar Phys.}}
 
\newcommand{\ssr}{    {\it Space Sci. Rev.}} 
 
\newcommand{\apjs}{    {\it The Astrophys. J. Suppl. Ser.}} 

\begin{document}

\begin{article}

\begin{opening}

\title{Trigger of successive filament eruptions observed by SDO and STEREO \\ }

\author{Sajal~\surname{Kumar Dhara}$^{1,2,a}$\sep
        B.~\surname{Ravindra}$^{3}$\sep
        Pankaj~\surname{Kumar}$^{4,5}$\sep
        Ravinder~\surname{Kumar Banyal}$^{3}$ \sep 
        Shibu~K.~\surname{Mathew}$^{1}$\sep      
        Bhuwan~\surname{Joshi}$^{1}$\sep     
       }
\runningauthor{Dhara~S.~K. et al.}

   \institute{$^{1}$ Udaipur Solar Observatory, Physical Research Laboratory, Dewali, Badi Road, Udaipur 313 004, India.\\
              $^{2}$ Istituto Ricerche Solari Locarno (IRSOL), Via Patocchi, CH-6605 Locarno-Monti, Switzerland.
                   $^{a}$  email: \url{sajal@prl.res.in}\\ 
              $^{3}$ Indian Institute of Astrophysics, II Block, Koramangla, Bangalore 560034, India.\\
              $^{4}$ Korea Astronomy and Space Science Institute (KASI), Daejeon, 305-348, Korea.\\
              $^{5}$ Heliophysics Science Division, NASA Goddard Space Flight Center, Greenbelt MD 20771, USA.\\
             }

\begin{abstract}
Using multi-wavelength observation from SDO and STEREO, we investigated the mechanism of two successive 
eruptions (F1 \& F2) of a filament
 in active region NOAA 11444 on 27 march, 2012. The filament was 
inverse `J' shaped and lying along a quasi-circular polarity inversion line (PIL). 
The first part of the filament  (F1) erupted at $\sim$2:30~UT on 27 March 2012, 
the second part of the filament (F2) erupted at around 4:20 UT on the same day.
A precursor/pre-flare brightening was observed below filament's main axis about 30 min 
prior to F1. The brightening was 
followed by a jet-like ejection below filament, which triggered the eruption. 
Before the eruption of F2, the filament
seems to be trapped within the overlying arcade
loops almost for $\sim$1.5~hr before its successful eruption. Interestingly, we 
observed simultaneously contraction ($\sim$12~km~s$^{-1}$) and expansion ($\sim$20~km~s$^{-1}$)
of arcade loops in the active region before F2. 
HMI magnetograms show the converging motion of the opposite polarities resulting in flux cancellation near PIL. 
We suggest that flux cancellation at PIL
resulted jet-like ejection below filament's main axis, which triggered the eruption F1 similar to tether-cutting process. 
The eruption F2 was triggered by removal of the overlying arcade loops via reconnection process.
Both filament eruptions produced high speed ($\sim$1000~km/s) CMEs.
\end{abstract}

\end{opening}

\section{Introduction}
     \label{S-Introduction}

Filaments/prominences are chromospheric and 
coronal structures, which contain relatively cool, dense plasma suspended along the polarity inversion 
line (PIL). Flares and CMEs are often associated with 
filament eruptions \citep{Schmieder02,Gopalswamy03}, 
which eventually affect space weather.
The fundamental processes responsible
for these eruptions 
originate in the magnetic fields of the solar atmosphere \citep{Priest02}.

According to flux rope model (e.g., \opencite{VanBallegooijen89}), the 
filament mass is supported in the dips of helical field lines. 
To understand the relation between the filament mass and corresponding supporting
magnetic structure, \inlinecite{Gilbert07} categorized filament eruptions observationally 
as ``full'', ``partial'' and ``failed'' eruptions, depending on the site of magnetic reconnection.
In case of a ``full eruption'', the reconnection takes place below the flux rope and most of the 
filament mass ($\geq$ 90\%) escapes along with the entire flux rope structure.
But for the ``failed eruption'', reconnection occur above the flux rope that 
suppress the filament downward.  In case of a ``partial eruption'', reconnection 
may occur within the  filament material that leads to the eruption of a portion of the filament. 

Several observations show that activity occurring near the filament is important 
during its pre-eruption phase.  Small-scale magnetic reconnection occurring during the pre-eruptive phase may 
trigger flares which involve large-scale reorganization of the magnetic connections (e.g., \opencite{Toriumi2013,Kusano2012}).
Therefore, it is very important to know the location and height of the small-scale reconnection sites. 
Several models proposed to examine the role of the initial magnetic reconnection which set up the conditions 
that are favorable for the magnetic core fields to erupt. One of the mechanisms is `tether cutting'
in which implosive/explosive reconnection occurs within the
twisted and deep sheared core field of a bipole above the polarity inversion line (PIL) of 
the arcade \citep{Moore92,Moore80,Moore01}. In this model the magnetic reconnection occurs 
in such away that it forms a flux rope as well as it triggers the eruption.
Initially filaments often undergo relatively slow rising motion before the onset of eruptions 
\citep{Kahler88,Liewer09,Nagashima07,Sterling07,Xu10,Sterling11}. These early slow rise motion 
was considered as one of the reason to trigger the eruption \citep{Moore06,Sterling07,Sterling10}. 
A transient brightening at or near the polarity inversion line (PIL),
coincident with emerging and/or canceling magnetic flux has been observed and are considered 
as precursors of the flare and filament eruption \citep{Martin80,Chifor07,Kim07,Liu09a, Sterling11}.

\cite{Antiochos99} proposed `breakout'  model, in which the topology of photospheric magnetic field is 
quadrupolar which plays a crucial role for the CME initiation.
The basic idea of this model is that magnetic reconnection at a null point in the corona above a sheared neutral line,
removes the constraint of the higher
magnetic flux tends to hold down the sheared low-lying field and thereby triggers
the sheared core field to erupt explosively outward \citep{Antiochos98,Antiochos99,Karpen12}.
In the context of 
both models, the role of magnetic reconnection for the
eruption can be investigated during the pre-eruption phase.

The magnetic reconnection can also occur at low level in the solar atmosphere 
(e.g., \opencite{Wang93}). It may be observed at the photosphere as a cancellation of magnetic 
features \citep{Priest94}. Flux cancellation \citep{Martin85} is the process 
where the magnetic flux disappears at the PIL; as formulated by
\inlinecite{VanBallegooijen89}. 
The flux cancellation at the polarity 
inversion line can lead to the coronal structure evolving towards highly sheared fields which eventually can erupt in the later
stage \citep{Green11}.
The gradual flux cancellation continuing over an 
extended period can trigger the filament eruption (e.g., \opencite{Sterling10,Zuccarello07}).

Filament eruptions occur in a very short time scale as compared
to the time scale of the coronal magnetic energy accumulation, hence there
should be MHD instability related to the trigger of the eruption. 
Therefore, different instability mechanisms have been evoked (review by \opencite{Forbes00}). 
Numerical MHD simulations of
the kink instability suggest that if the twist of the flux rope exceeds
a critical value, then flux rope becomes
unstable (e.g., \opencite{Hood79}, \opencite{Torok03}, \opencite{Torok04}). This value
depends on the aspect ratio of the loop, the ratio of the plasma to magnetic
pressure. Previous observational studies support the kink instability to be the triggering mechanism 
for the solar eruptions (e.g., \opencite{Srivastava10}, \opencite{Kumar2012}, \opencite{Kumar2014}). 
One of the example of a filament eruption studied by \cite{Yan14} for the active region NOAA 11485 
showed that the leg of the filament rotated up to 2.83$\pi$ around the axis of 
the filament with a maximum rotation speed 100 degrees/minute. This study indicates that the kink
instability is the trigger mechanism for the solar filament eruption.

The filament/flux rope can experience an another type of instability called ``torus
instability''(e.g., \opencite{Kliem06}, \opencite{Zuccarello14}, \opencite{Dhara14}). 
Some pre-eruptive processes, such as flux cancellation/emergence at the neutral line and 
magnetic reconnection during the observed
brightening, initiate the eruption by bringing the flux rope to a height of torus instability.
Using MHD numerical simulations in a 3D spherical geometry of emergence of a flux rope from
the subsurface into the magnetized corona, \cite{Fan07} finds that when the background
magnetic field decreases slowly with height, a strongly-twisted flux tube emerging out of the solar
surface can rupture through the arcade field via kink instability. But when the
background magnetic field decreases rapidly with height, then the flux rope can become unstable to the lateral 
expansion and then it erupts. This kind of loss of equilibrium could be interpreted as torus instability
\citep{Chen11}. In this model, a current ring of major radius
$R$ is embedded in an external magnetic field ($B_{ex}$). 
The current rings are subjected to a 
radially outward-directed hoop force. When the ring expands, the hoop force decreases. 
If the inward-directed Lorentz force due to the external field ($B_{ex}$) decreases faster with radius $R$ than the
hoop force, the system is unstable to perturbations. The decay index for a current ring of major radius $R$ and
embedded in an external field $B_{ex}$, is defined as follows \citep{Zuccarello14}:
 \begin{equation}
 n= -R\frac{d}{dR}(ln B_{ex})
 \end{equation}
Assuming external magnetic field $B_{ex}$ $\varpropto$ $R^{-n}$, \cite{Bateman78} numerically showed such an instability
will occur when $ n$ $>$ $n_{critical}$ = 1.5.

In this paper, we study the origin of two successive filament eruptions (F1 \& F2) that occurred in AR
NOAA 11444, observed by SDO and STEREO. 
 Here we present the important morphological
changes in and around the filament prior to eruptions and the possible triggering mechanism
of the filament eruptions (F1 and F2). 
This paper is organized as follows. In the next section, we describe about the different 
instruments and data used in this study. A brief description about filament and a sequences of the events 
in the corona, chromosphere as well as at the photosphere are 
presented in Section~\ref{section3}. Finally, we discuss our results
in Section~\ref{section4}.    
     
  \section{Instruments and Data} 
      \label{section2}   
      
We used the EUV images obtained from \textit{Atmospheric Imaging Assembly} ({AIA;} \opencite{Lemen12}) 
on board the \textit{Solar Dynamics Observatory} ({SDO;} \opencite{Pesnell12}) to study the filament 
eruption in detail. In particular we have used data obtained in AIA 171, 193 and 304~\AA~
channels which correspond to the corona and chromosphere. 
We have used Level-1.0 data. 
We used ${\it{aia\_prep.pro}}$ routine available in SSW packages to co-align the images 
from all of the AIA channels (171, 193 and 304~\AA). 
From these data set the filament regions are extracted and tracked over time.
These images are obtained with a cadence of 12~seconds and
a pixel resolution of 0.6$^{\prime\prime}$. We acquired the data starting from 22:00~UT 
on March 26, 2012 until 07:00~UT on March 27, 2012 which covers activation of the filament 
eruption and subsequent runaway motion for the entire event. 
To study the filament eruption above the solar limb, we also used EUVI 304 and 195~\AA~images 
from the Sun-Earth Connection Coronal and Heliospheric Investigation ({SECCHI;} \opencite{Howard08})
onboard the Solar-Terrestrial Relations Observatory-A ({STEREO;} \opencite{Kaiser08}).
The EUVI's 2048$\times$2048 pixel detectors have a field of view out to 1.7 solar radii, which completely covered
this filament eruption event.

The morphology of the filament at chromospheric heights can be studied with H$\alpha$ 
images. We obtained full-disk H$\alpha$ images from \textit{Global Oscillation Network Group} (GONG; \opencite{Harvey96}) 
which collects H$\alpha$ images at six sites around the world. The dark, flat, smear corrected and
compressed H$\alpha$~images are available with almost 1-minute cadence. These H$\alpha$~images are obtained with 
2k$\times$2k pixel CCD camera whose pixel resolution is about 1$^{\prime\prime}$. We acquired 
H$\alpha$ images starting from 23:00~UT on March 26, 2012 until 06:00~UT on March 27, 2012. The data set 
covers first and second eruption of the filament and subsequent flares.

In order to examine the changes in the photospheric magnetic field near the filament 
footpoints, we utilized the full-disk line-of-sight magnetograms at level-1.0
from \textit{Helioseismic and Magnetic Imager} ({HMI;} \opencite{Scherrer12}) 
with a cadence of 45~seconds. The line-of-sight magnetogram has pixel resolution of 
0.5$^{\prime\prime}$. We used ${\it{aia\_prep.pro}}$ routine available in SSW packages to 
upgrade HMI magnetogram to level-1.5 from level-1.0 so that it is interpolated to the AIA 
pixel-resolution. We have tracked the ROI by utilizing the heliographic co-ordinate information 
and finally corrected for the line-of-sight effect by multiplying 1/cos$\theta$, where $\theta$ 
is the heliocentric angle. We then averaged 4 magnetograms to reduce the noise level \citep{Liu2012} to about 10~G.
These magnetograms have been used to study the evolution of magnetic 
fields in and around the filament at the photospheric level.

\section{Observations and Results} 
  \label{section3}

\subsection{AIA observations of the filament eruption}

\begin{figure}   
 \begin{center} 
 
  \centerline{\hspace*{0.02\textwidth}
               \includegraphics[width=0.58\textwidth,clip=]{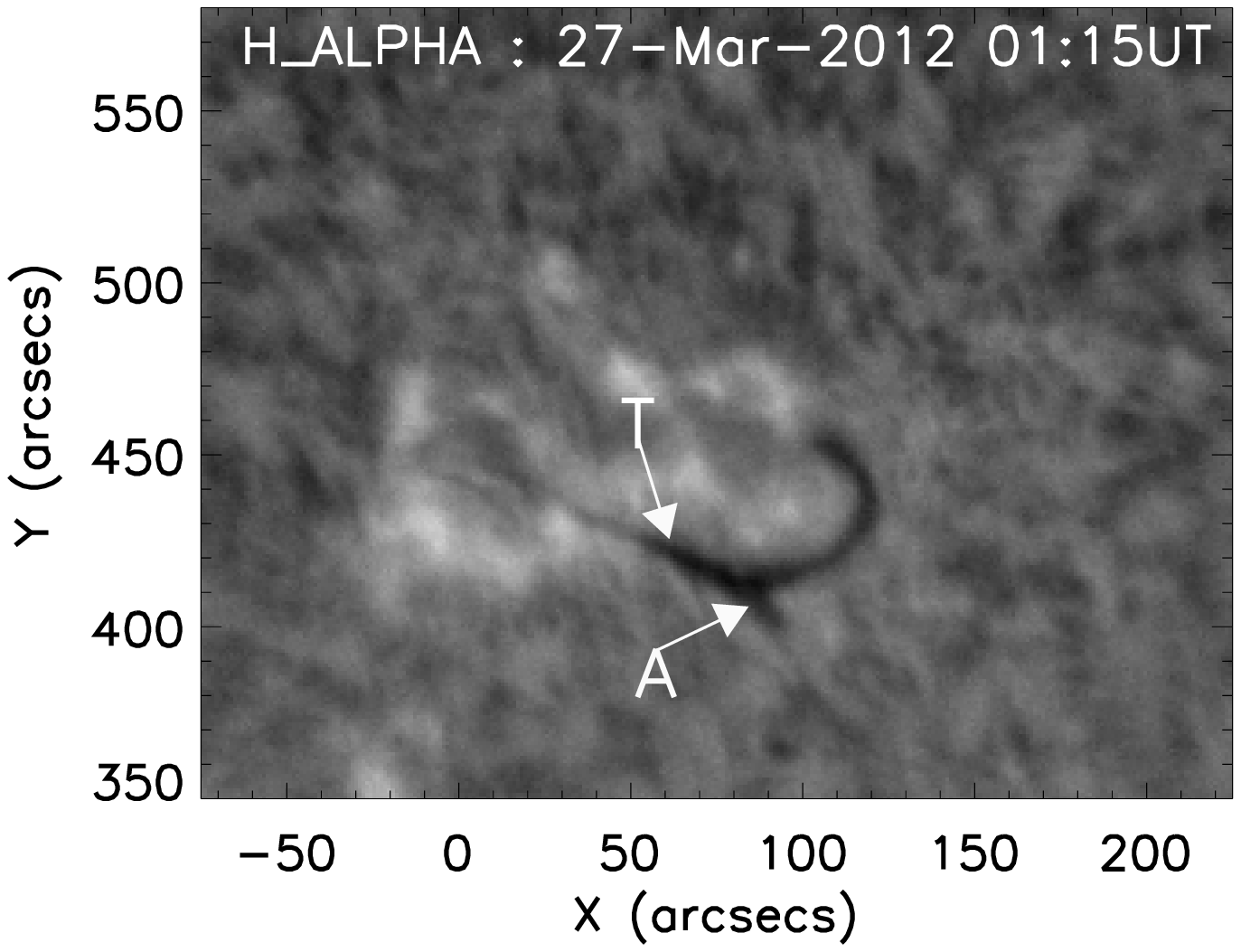}
               \hspace*{-0.14\textwidth}
               \includegraphics[width=0.58\textwidth,clip=]{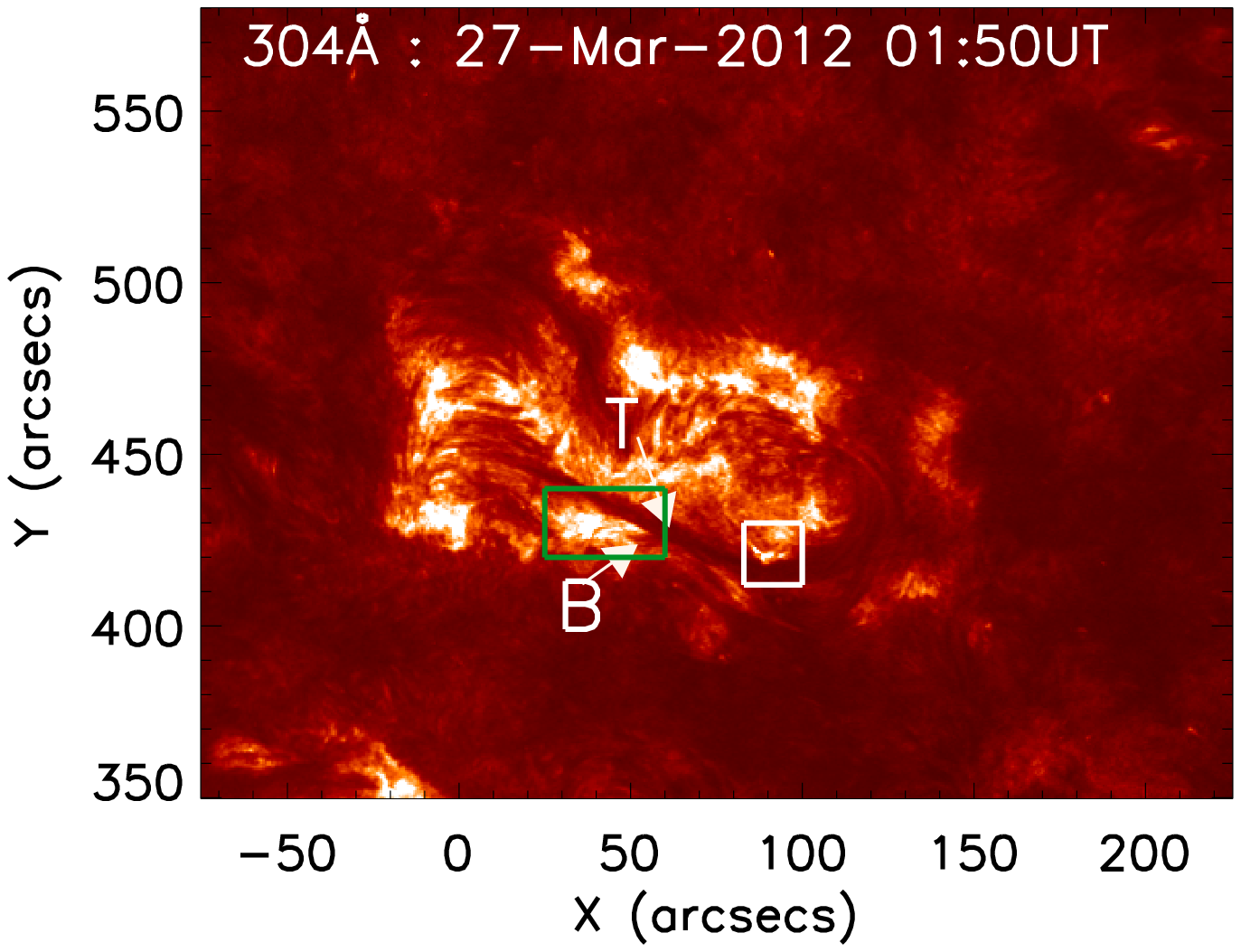}
              }
              \vspace{-0.05\textwidth}   
              
   \centerline{\hspace*{0.02\textwidth}
               \includegraphics[width=0.58\textwidth,clip=]{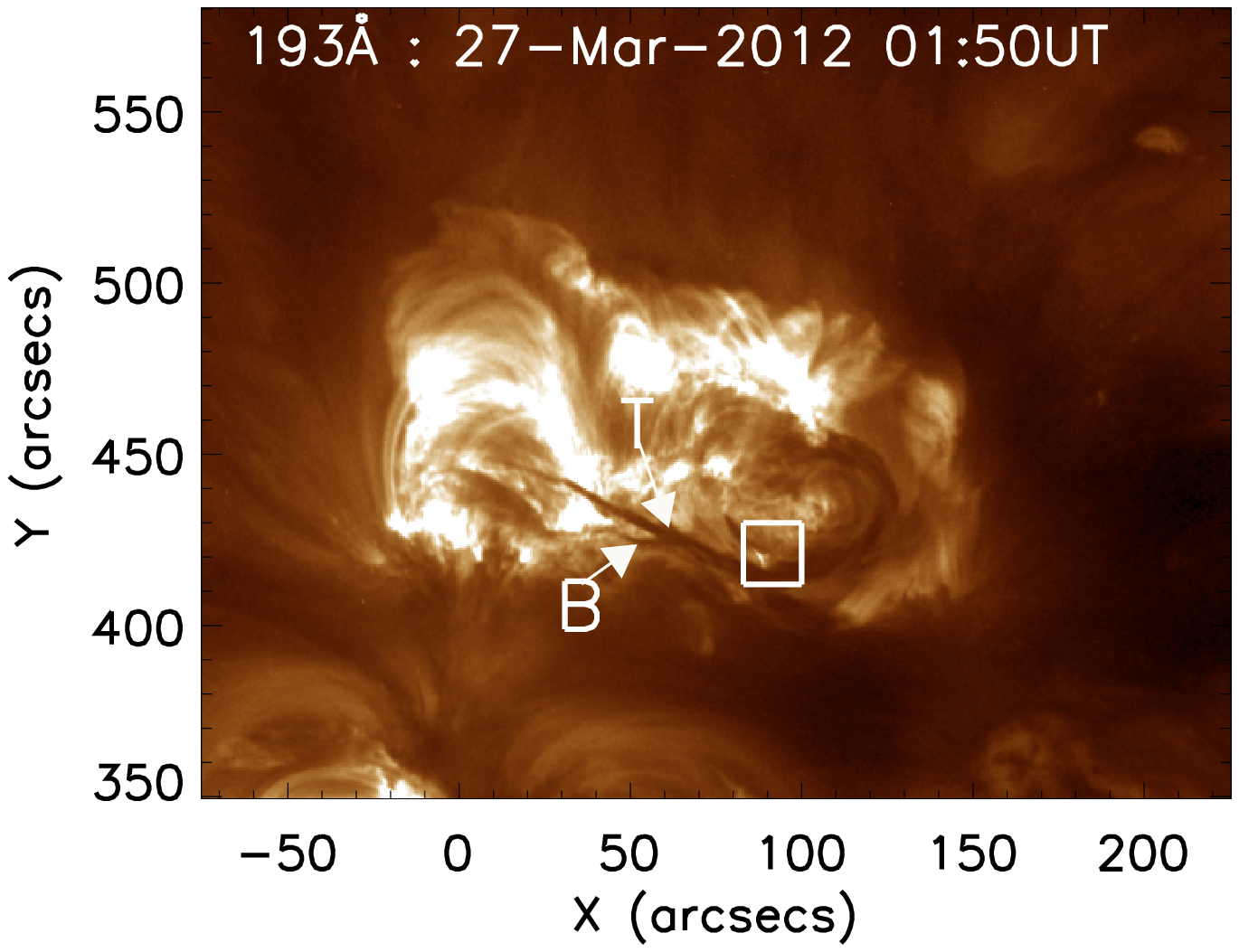}
               \hspace*{-0.14\textwidth}
               \includegraphics[width=0.58\textwidth,clip=]{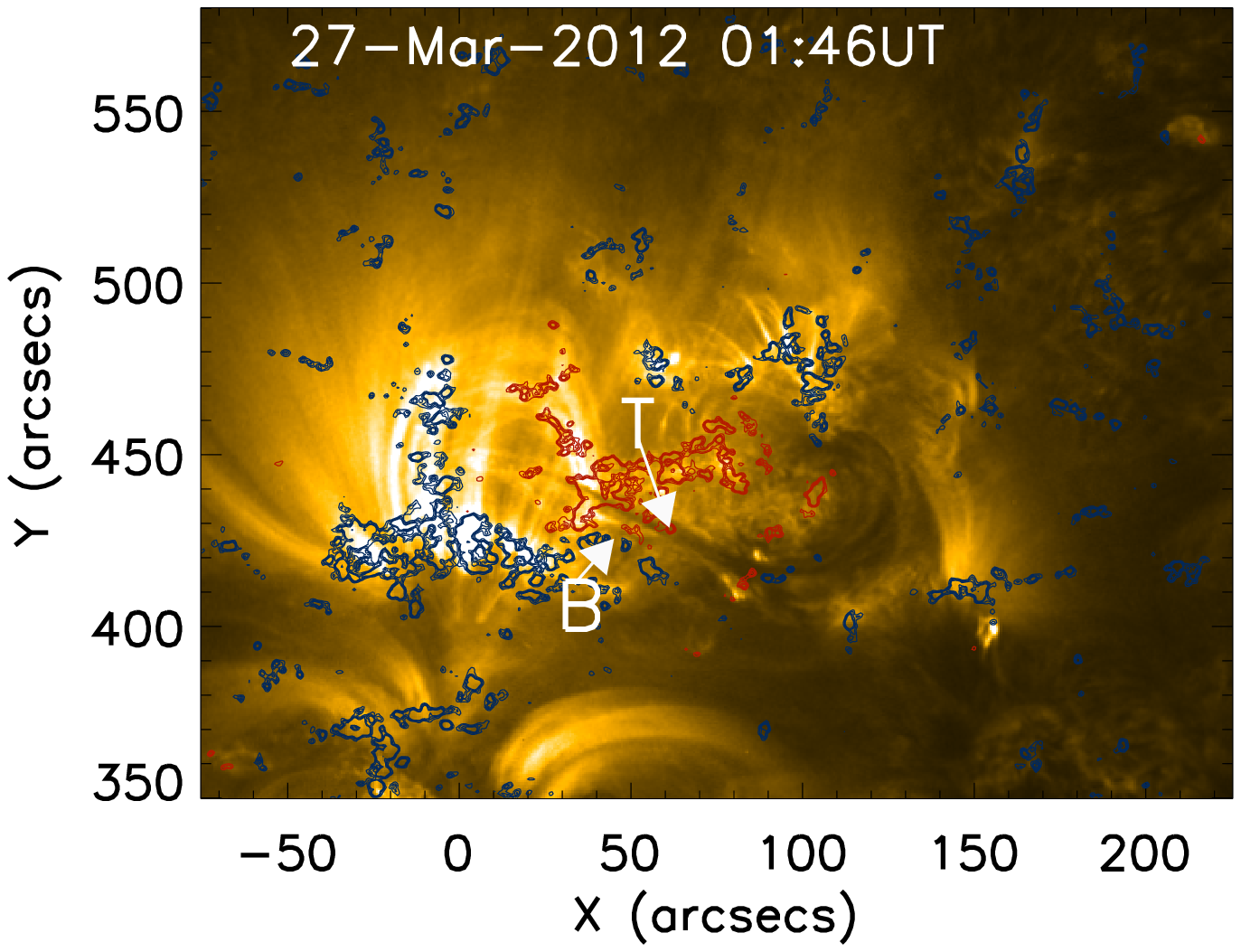}
              }
     \vspace{-0.05\textwidth}   
    
  \end{center}     
\caption{Filament observed in H$\alpha$ (top-left), AIA
304~\AA~(top-right), 193~\AA~(bottom-left) and 171~\AA~(bottom-right).
The filament location is shown by white arrow with `T' symbol. The arrow
with `B' symbol shows a bifurcation of the filament below T. The arrow
with `A' symbol in the H$\alpha$ image shows the location of the protrusion which appears as barb of the filament.
The white boxed region in 304 and 193~\AA~images shows the jet location close to the filament.
The green boxed region in the top-right panel shows the location of the preflare brightening in 304~\AA~image. 
The contours of the magnetic 
fields are overlaid on the 171~\AA~image. The red and blue contours represent the 
 positive and negative polarities with magnetic field strength values of $\pm$ 100, 150, 200 
and 250~G, respectively.}

\label{fig:ch1_1}
\end{figure}

\begin{figure}   
\begin{center}  
  \centerline{\hspace*{0.015\textwidth}
               \includegraphics[width=0.8\textwidth,clip=]{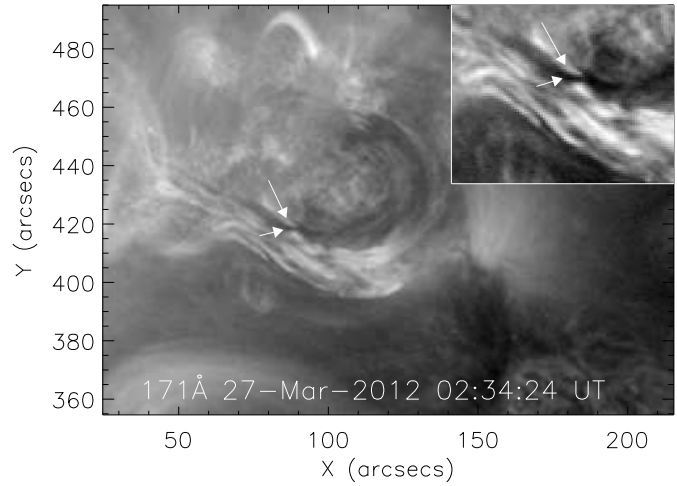}
             }
     \vspace{-0.05\textwidth}    
\end{center}        
\caption{ Erupting filament shown in 171~\AA~ channel. The two arrows
indicate crossing of dark threads over the bright threads. The box on the top
right of the image shows the zoomed in version of the cross over threads.}
\label{fig:171_hel_img}
\end{figure}

The filament was observed in the active region (AR) NOAA 11444 
at the location of N19 E26 on March 23, 2012 and survived for almost 3~days. On March 27, 2012, before the eruption, the 
filament was observed at heliographic position of N21~W17. 
The filament appeared as inverse {\it J} shaped in  H$\alpha$ (top-left),
He~II~304 (top-right), 193~\AA~(bottom-left) and 171~\AA~(bottom-right) images  in Figure~\ref{fig:ch1_1},
respectively. The arrow followed by letter `T' in Figure~\ref{fig:ch1_1} show the filament 
and the arrow followed by letter `A' in
H$\alpha$ image indicates the location of the protrusion which appears as barb. 
 The appearance of the filament in the 
coronal images (193 and 171~\AA) are also shown in Figure~\ref{fig:ch1_1}~(bottom-left and right, respectively). 
The filament followed the polarity inversion line. The northern end of the filament rooted in compact negative polarity plage region 
and the southern portion ended in the positive polarity plage. The axial field component of 
the filament is right bearing when we look at the filament from positive polarity side. 
This corresponds to the dextral chirality and is consistent with hemispheric helicity rule \citep{Martin98,Martin1994}.
It was confirmed further in erupting filament. 
 The erupting filament (shown in Figure~\ref{fig:171_hel_img}) shows a crossing of dark and
bright threads in 171~\AA~image. The crossing of dark features over
the bright regions is easily identified in the zoomed image, where a
bright feature in the background is seen going from left to right while
the dark feature in the foreground crosses it from right to left. This
crossing threads correspond to the negative mutual (type III) helicity \citep{Chae2000}.

The filament seems to have two structures.
One is broad inverse {\it J} shaped which is
visible in H$\alpha$--6563~\AA, AIA~171, 193 and 304~\AA~wavelengths. This filament region is
shown by an arrow with letter `T' symbol, indicating the top most filament. One more thin 
filament lying below the top filament or it could be just a bifurcation of the main filament, shown 
by an arrow with `B' symbol. 
It appears to start at a location of 
$\sim$35$^{\prime\prime}$ in the horizontal direction and at $\sim$425$^{\prime\prime}$ in the
vertical direction in EUV images. It bends down at $\sim$45$^{\prime\prime}$ and later continues with the main `T' 
filament, also taking inverse J shape.
The bottom side or bifurcated filament is shown by the letter `B' in Figures~\ref{fig:ch1_1}.


In order to study the filament eruption in detail, we created an online movie\footnotemark[1] of the erupting filament 
in 304 and 171~\AA~channel of AIA. The movie clearly shows that the 
filament erupted in two stages. In the first stage, the southern part of the filament erupted.
This eruption is followed by a C5.3~class flare identified by GOES detector at $\sim$02:52~UT.  
The second eruption initiated at around 3:50~UT during which the northern part of the filament erupted
and accelerated at 4:20~UT followed by a C1.7~class flare at 
$\sim$04:25~UT. 

\footnotetext[1]{Movies (AIA$\_$304.avi and AIA$\_$171.avi) generated from SDO/AIA 171
and 304~\AA~images, respectively, of the filament 
eruption associated with active region NOAA 11444 are available in the electronic version of the manuscript.}


\begin{figure}   
\begin{center}  
  \centerline{\hspace*{0.01\textwidth}
               \includegraphics[width=1.08\textwidth,clip=]{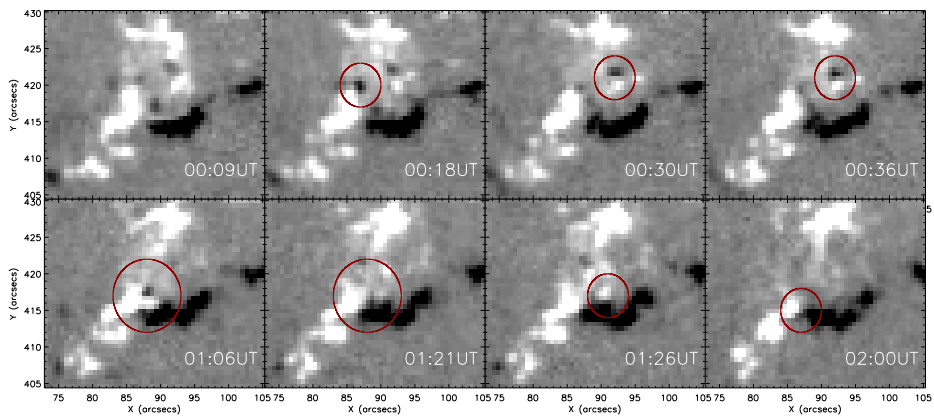}
             }
     \vspace{-0.05\textwidth}    
\end{center}        
\caption{A sequence of line-of-sight magnetograms showing the location of the jet, seen in 
the white boxed region in Figure~\ref{fig:ch1_1}.
The white and black region represent the positive and negative polarities, respectively. The 
locations of the flux cancellations and emergence are shown by circles.}
\label{fig:ch1_3}
\end{figure} 


The contours of photospheric magnetic fields overlaid upon the 171~\AA~images, shown in Figure~\ref{fig:ch1_1} (bottom-right), 
indicate that the filament is lying along 
the neutral line and the southern portion of the filament ending in the
bipolar regions where the flux cancellation was observed. The jet in 
the cusp shaped regions observed close to the filament is shown by white boxed region in 
the 304 and 193~\AA~wavelengths in Figure~\ref{fig:ch1_1}. Over the period of 
2.5~hrs starting from 00:00~UT to 02:30~UT (on March 27, 2012), jets are repeatedly observed in this region.
The photospheric magnetograms showed repeated emergence and 
cancellations of magnetic flux at the same site. Figure~\ref{fig:ch1_3} shows the sequence of line-of-sight magnetograms 
for the same location at different times during the observations of jet. The emergence and cancellation of magnetic flux
regions are marked by circles in the time sequence of magnetograms.


\subsection{Pre-flare brightening and filament eruptions (F1 \& F2)}

\begin{figure}    
 \begin{center} 
  \centerline{\hspace*{-0.4\textwidth}
               \includegraphics[width=1.4\textwidth,clip=]{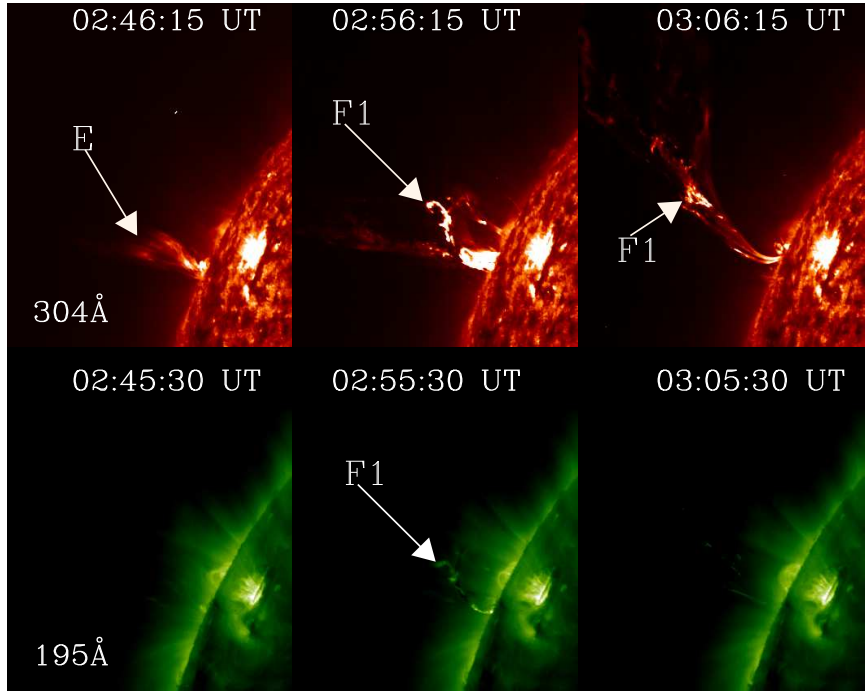}
              }
              \vspace{-0.15\textwidth}   
  \end{center}     
\caption{Sequence of STEREO-A/EUVI 304~\AA (top panel) and 195~\AA~ (bottom panel) 
images show the eruption of filament (F1). The arrow indicated by `E' in 304~\AA~ image shows the
upward plasma flow/jet before the eruption of F1.}
\label{fig:stereo_1}
\end{figure}

A brightening was observed starting at 01:56~UT below the
filament T. The brightening location
is shown in 304~\AA~channel by a green box in Figure~\ref{fig:ch1_1}~(top-right).
The brightening was followed by a bright plasma flow which moved from east end of the 
filament to the west. During the flow, the filament appeared as a sequence of bright and 
dark threads.  The bright flow reached the barb location at 02:03~UT. 
During that time the bifurcated filament (B) moved up a little and
also raised the top filament (T). 
At around 2:35~UT the separation between the bright and dark threads of the filament became
large and in the west side, one end of the filament started to move upward. 
At around 2:53~UT there was a 
C5.3 class flare and during that time a large amount of bright mass was ejected along with the 
filament eruption (F1). 
Figure~\ref{fig:stereo_1} shows the sequence of STEREO-A/EUVI 304~\AA~and 195~\AA~images during the eruption of F1.
The first panel of Figure 4 (top) shows the bright upward flow/jet which is seen in 304 images before the eruption.
The middle and last panel shows the
eruption of the filament. 
The estimated speed of filament eruption (F1), computed from 304~\AA~ images, is 274 $\pm$ 11~km/s.
The formation of a close underlying flare loop, observed in the 195~\AA~ image during the jet, suggests 
the magnetic reconnection as a driver of jet and associated filament eruption.

Soon after the eruption of F1, we observed the activation of filament F2 that does not erupt and 
remains stationary for a longer period.
This filament (F2) is shown by an arrow in 171~\AA~image in Figure~\ref{fig:ch1_8} (top). 
At around 4:20~UT this filament also started to erupt 
followed by a C1.7 class flare. The filament disappeared from the 
field-of-view at around 4:30~UT in the AIA 171~\AA~images.   


\subsection{Temporal evolution of filament eruptions and flares}

\begin{figure}     
\begin{center}  
             \vspace{-0.12\textwidth}
  \centerline{\hspace*{0.02\textwidth}
               \includegraphics[width=0.85\textwidth,clip=]{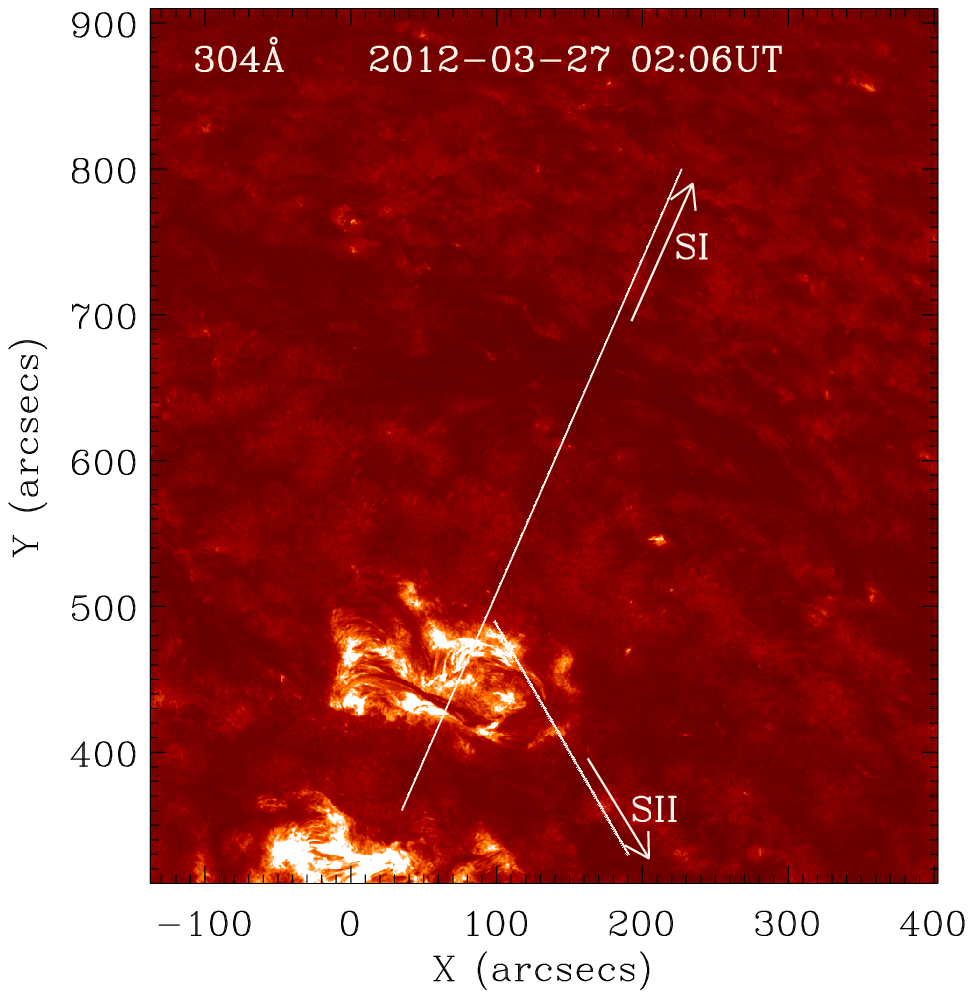}
              }
              \vspace{-0.02\textwidth}   
      
   \centerline{\hspace*{0.01\textwidth}
               \includegraphics[width=0.52\textwidth,clip=]{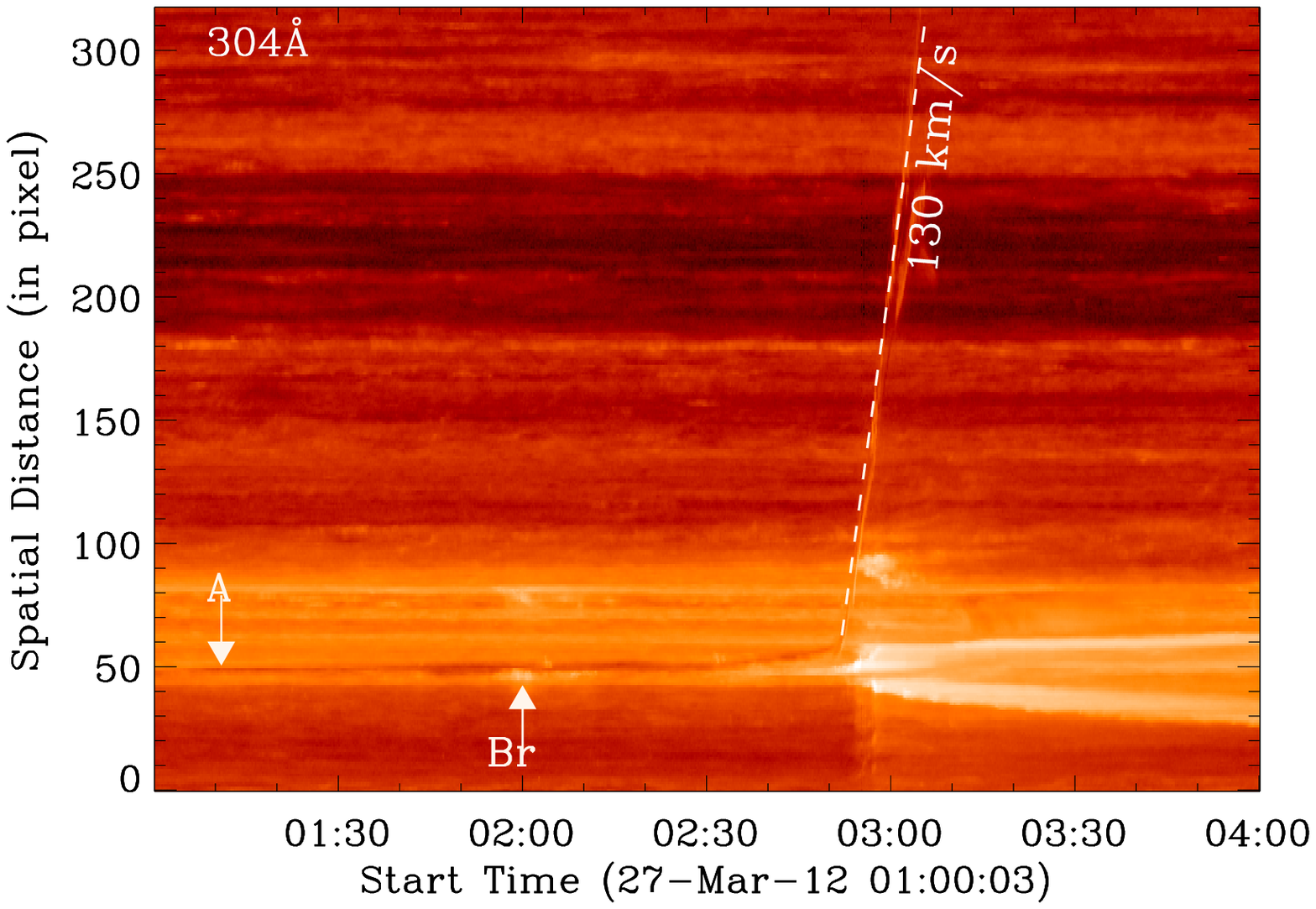}
               \hspace*{-0.05\textwidth}
               \includegraphics[width=0.52\textwidth,clip=]{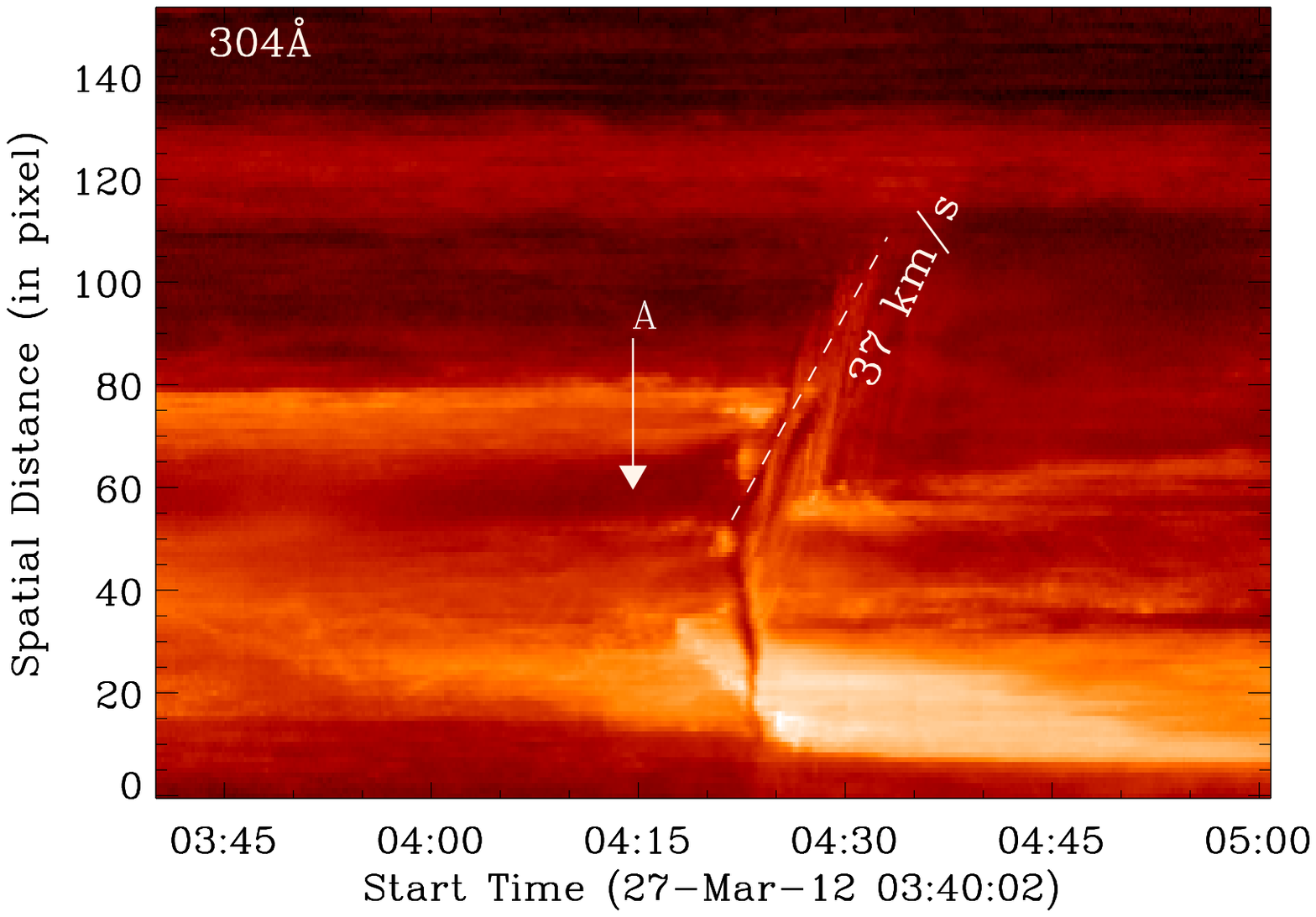}
              }
     \vspace{-0.27\textwidth}    
    
    \centerline{\Large \bf     
      \hspace{0.0 \textwidth}  
      \hspace{0.415\textwidth}  
         \hfill}
     \vspace{0.22\textwidth}    
   \centerline{\hspace*{0.01\textwidth}
               \includegraphics[width=0.52\textwidth,clip=]{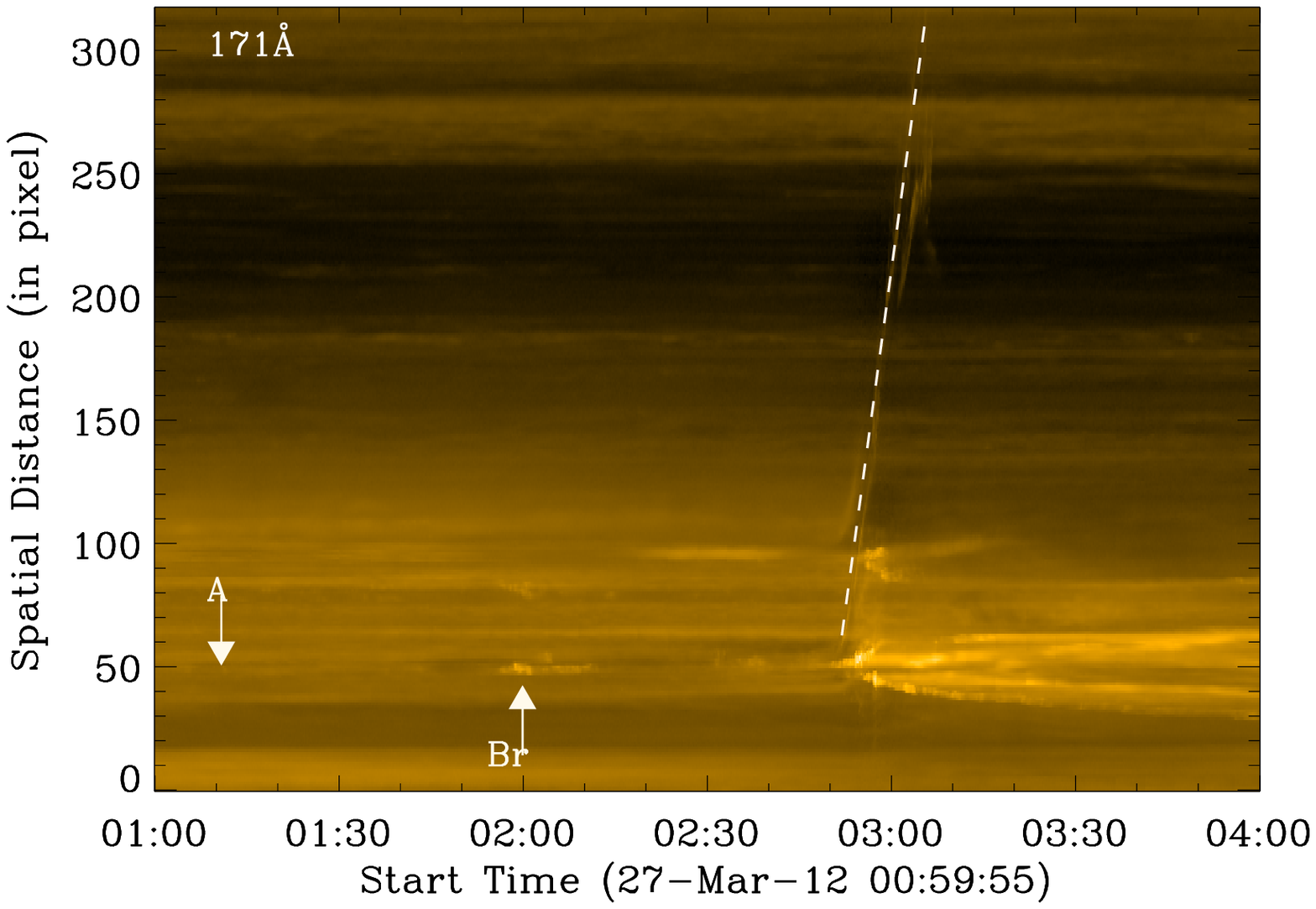}
               \hspace*{-0.05\textwidth}
               \includegraphics[width=0.52\textwidth,clip=]{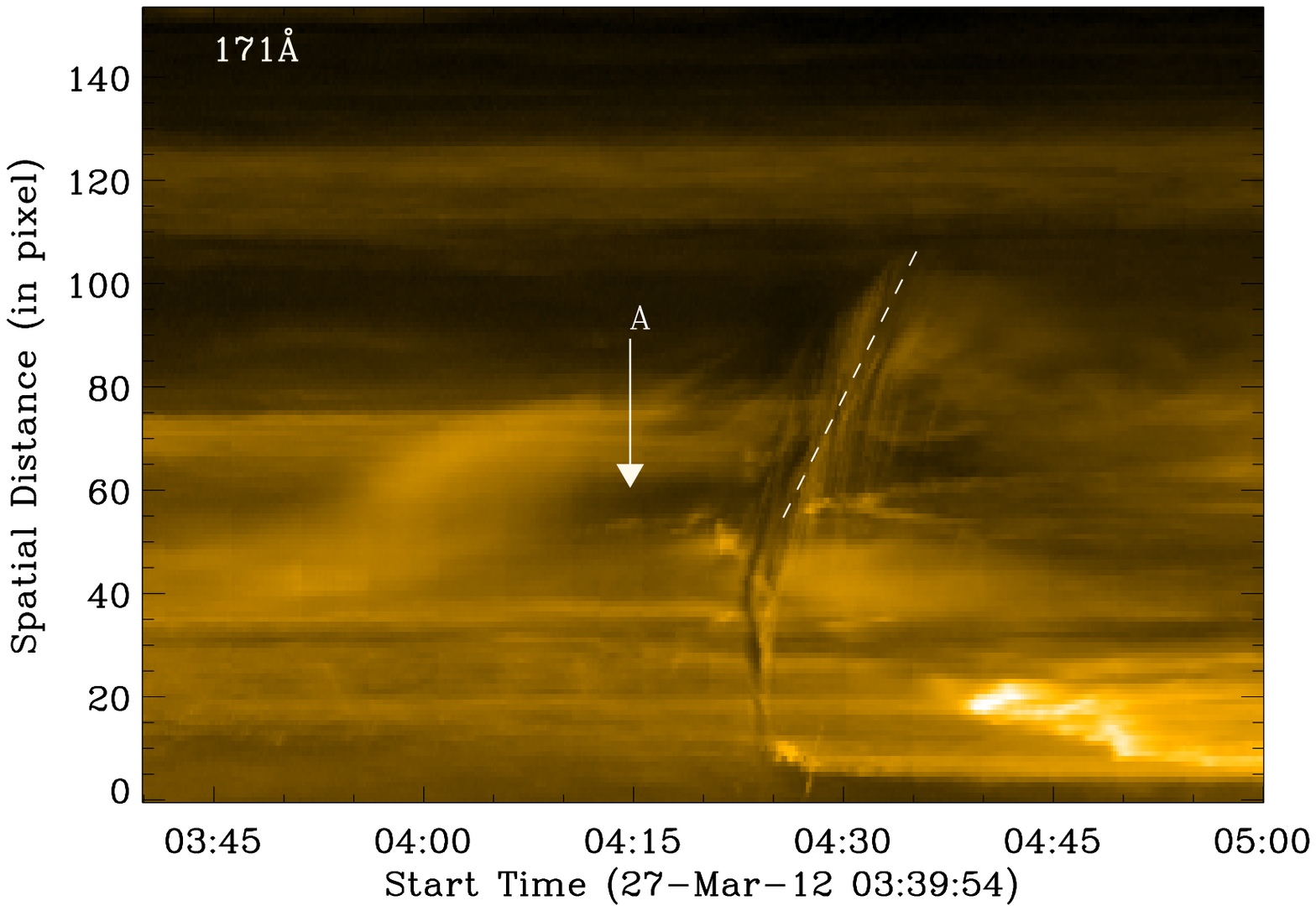}
              }
     \vspace{-0.05\textwidth}    
   
\end{center}        
\caption{Top: AIA 304 \AA~ image showing the filament. Two slits overlaid on the 
image represent the positions to generate the time-distance maps. The
slit-I crosses 1$^{st}$ eruption of the filament (F1). The slit-II crosses 2$^{nd}$ eruption of 
the filament (F2). Middle: The space--time map of filament
eruption for slit-I (left) and for slit-II (right) respectively extracted from 304~\AA~images. Bottom: The 
space--time map of filament eruption for slit-I (left) and for slit-II (right) respectively 
extracted from 171~\AA~images. A dark bent 
portion of the filament represents the path of the filament while erupting. 
The white arrow followed by letter `A' (left images) shows initial position of the filament before eruption.
Other white arrow followed by letter `Br' shows the brightening during the eruption.}
\label{fig:ch1_5}
\end{figure}

To investigate the temporal evolution of the successive filament eruptions and associated flare activities, we
constructed TD maps from a series of AIA 304 and 171~\AA~images starting from  01:00~UT on March 27, 2012. The 
TD map for the first filament eruption is extracted from slit-I and the second eruption is extracted 
from slit-II (shown in Figure~\ref{fig:ch1_5}~(top)). The TD maps shown in 
Figure~\ref{fig:ch1_5} are extracted from 304 and 171~\AA~images. The position of the filament in 
TD map generated from 304~\AA~images (middle-left image), is shown by an arrow `A'. 
There was a pre-flare brightening close to the filament at $\sim$01:56~UT. The location of the brightening 
in TD map is shown by an arrow `Br'. A similar brightening is also visible in other TD map, 
generated from 171~\AA~images (bottom-left). Later, the filament showed
slow-rise motion at a projected speed of 1.5$\pm$0.3 km~s$^{-1}$. 
At around 2:30~UT, there was another brightening which is related to the C5.3 class flare. The filament
rose faster during this time. This location is
indicated by a vertical line close to the filament in TD map. The flare maximum 
occurred at around 3:08~UT. The filament accelerated during the flare and erupted with a 
projected speed of 130$\pm$3~km/sec.

\begin{figure}     
 \begin{center} 
  \centerline{\hspace*{0.22\textwidth}
               \includegraphics[width=1.4\textwidth,clip=]{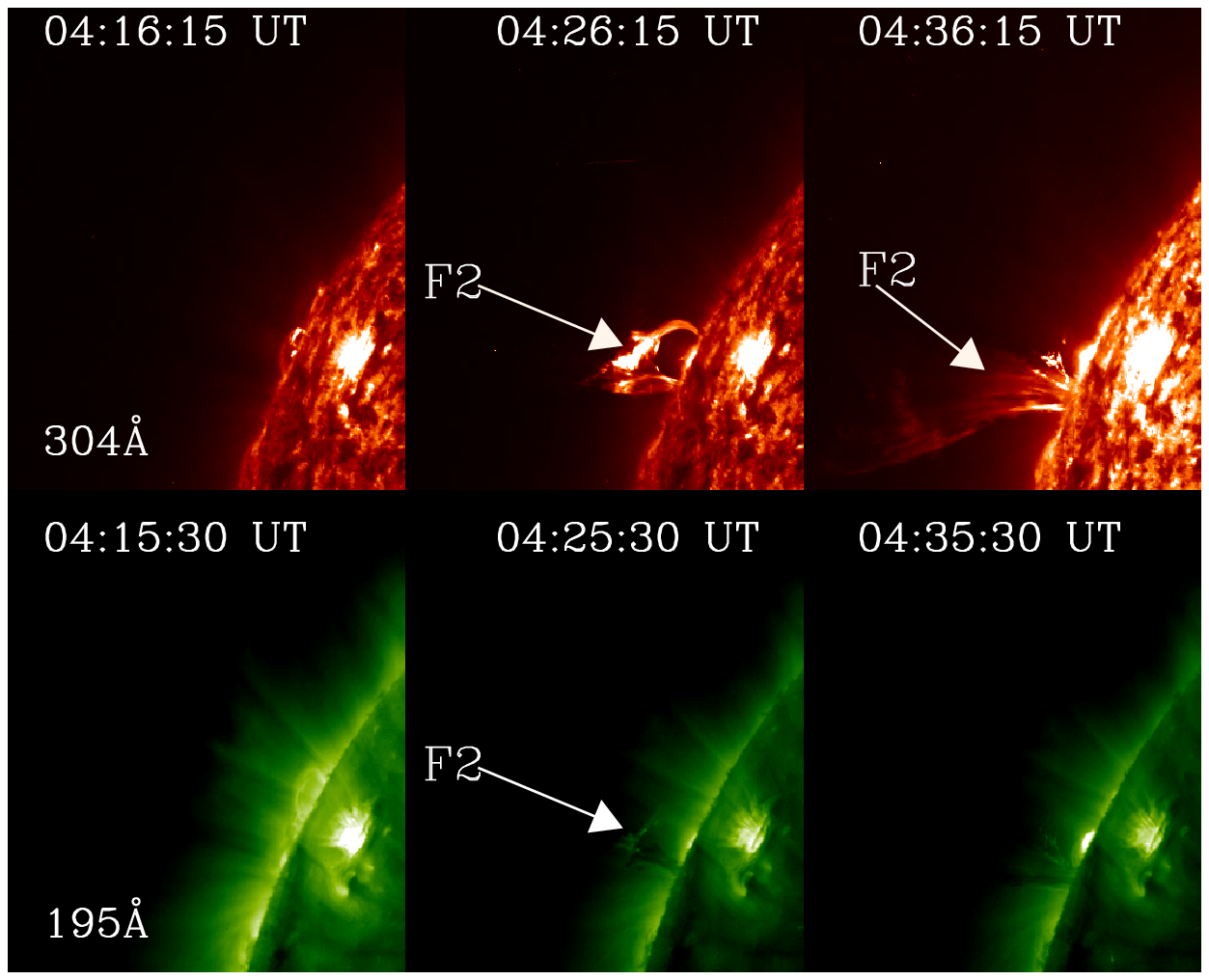}
              }
              \vspace{-0.18\textwidth}   
  \end{center}     
\caption{Sequence of STEREO-A/EUVI 304~\AA~ (top panel) and 195~\AA~ (bottom panel) 
images showing the eruption of filament (F2).}
\label{fig:stereo_2}
\end{figure}

The second filament F2 erupted in a different manner compared to the previous eruption. After
its activation, filament seems to have a counter clockwise motion and
then moved away towards the south. This filament accelerated and erupted with a projected
speed of 37$\pm$2~km/sec. This event was  followed by the C1.7 class flare.
 Figure~\ref{fig:stereo_2} shows the sequence of 
STEREO-A/EUVI 304~\AA~and 195~\AA~images during eruption of the filament (F2) in second phase. 
The speed during eruption of F2 on the sky plane estimated from these 304~\AA~images was about 145$\pm$10~km/sec.


\begin{figure}     
\begin{center}  
  \centerline{\hspace*{-0.01\textwidth}
               \includegraphics[width=0.8\textwidth,clip=]{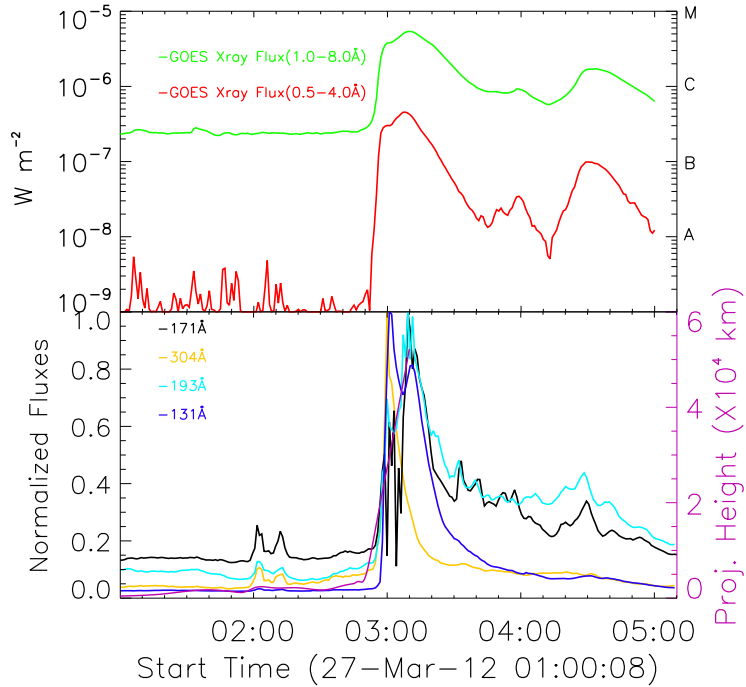}
              }
     \vspace{-0.05\textwidth}    
\end{center}        
\caption{
Top: GOES soft X-ray flux profiles in the 1-8 and 0.5-4~\AA~channels. Bottom: The normalized 
fluxes are plotted for the AIA 171~\AA, 193~\AA, 304~\AA, and 131~\AA~ channels. The projected 
time-distance plot for the eruption F1 is also included at the right y-axis.}
\label{fig:ch1_6}
\end{figure}

Figure~\ref{fig:ch1_6} (bottom panel) shows the intensity profiles in the 
AIA 171, 304, 131 and 193~\AA~channels. The pre-flare brightening was observed below the 
 filament. The light-curve is obtained by 
integrating the intensity in a small region (green colored box) shown in the 
AIA 304~\AA~image (Figure~\ref{fig:ch1_1} (top-right)).  
The light-curves showed a small rise 
at around 01:56~UT in the boxed region. The TD maps indicated that the filament
activation started at the same time. Later, at around 2:50~UT there was a jump in the GOES
X-ray curve in the 1.0--8.0~\AA~band. At the same time we observed a jump in intensity for all the 
304, 131, 171, and 193~\AA~wavelength bands.
Therefore, the EUV and X-ray flux profiles are consistent. 
The brightening in the GOES and in other EUV wavelengths reached the background level 
over a period of about 40~minutes after the flare. During the initiation of the C5.3 
flare, the filament showed an acceleration and eventual eruption.
Interestingly, GOES soft X-ray flux profiles reveal a small B-class flare (peak around 03:56 UT) 
between these two filament eruption. However, it can not be distinguished in the EUV light curves. This flare was 
associated with the eruption of an overlying loop arcade prior to the successful eruption of filament F2.

\subsection{Contraction of a coronal loop in the first phase}

\begin{figure}     
\begin{center}  
  \centerline{\hspace*{0.03\textwidth}
               \includegraphics[width=1.12\textwidth,clip=]{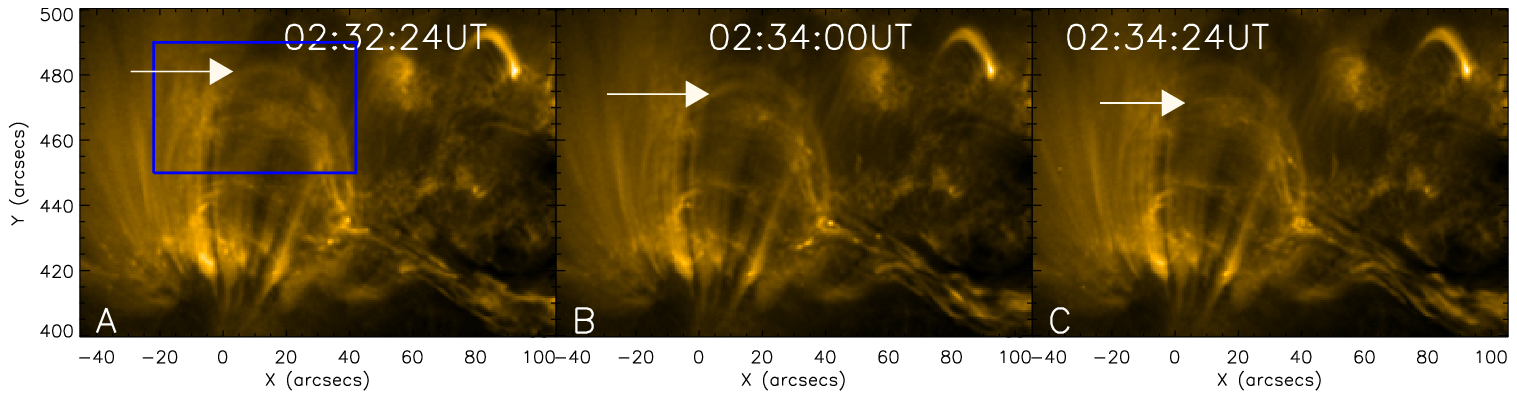}
              }
              \vspace{-0.03\textwidth}   
   \centerline{\hspace*{0.03\textwidth}
               \includegraphics[width=1.12\textwidth,clip=]{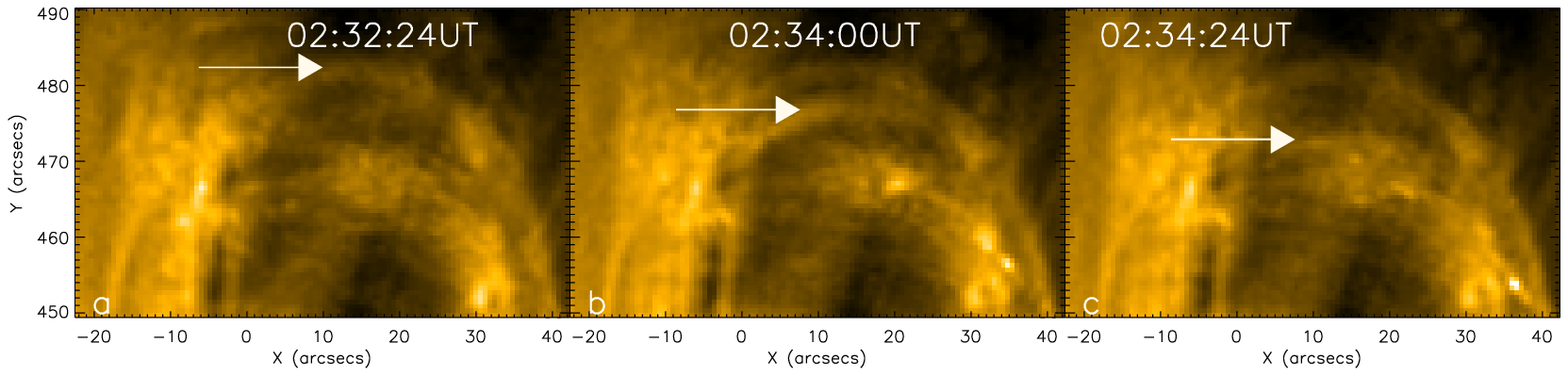}
              }                                   
              \vspace{0.02\textwidth}   
   \centerline{\hspace*{0.03\textwidth}
               \includegraphics[width=1.12\textwidth,clip=]{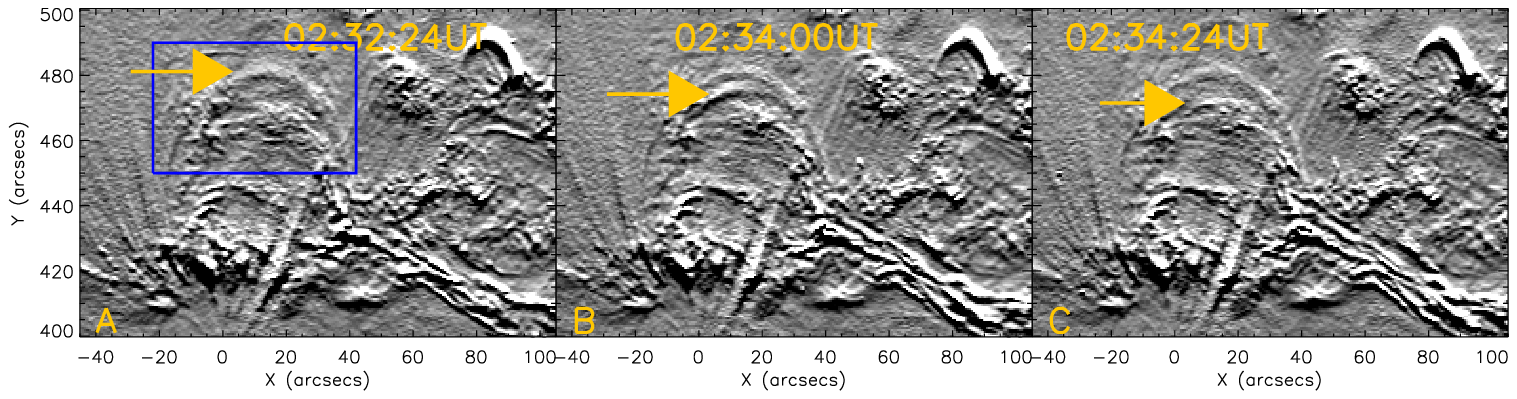}
              }
              \vspace{-0.03\textwidth}   
   \centerline{\hspace*{0.03\textwidth}
               \includegraphics[width=1.12\textwidth,clip=]{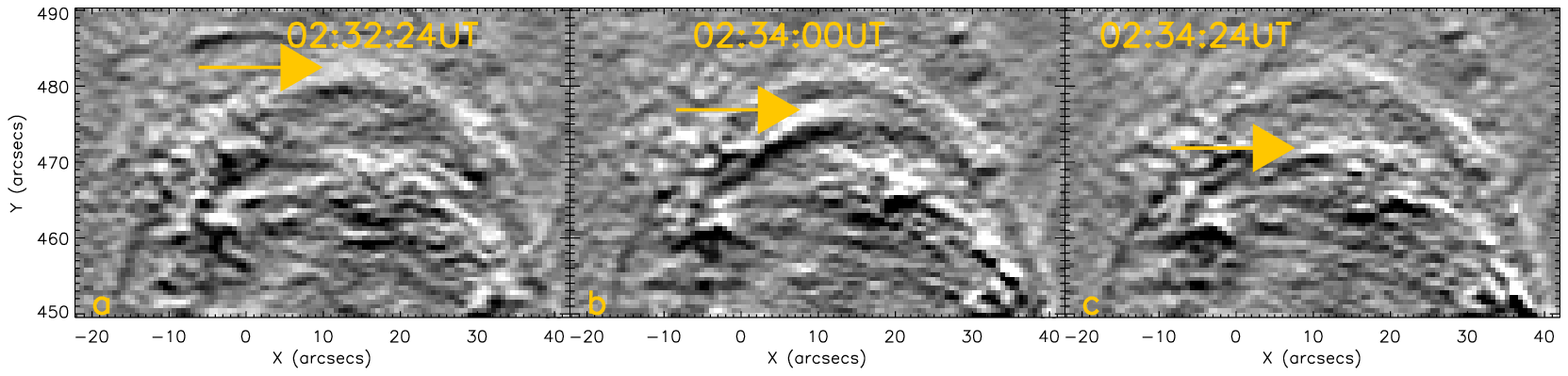}
              }
     \vspace{-0.05\textwidth}    
\end{center}        
\caption{Top -- 1st panel: Sequence of 171~\AA~images (A, B, C) for 
the east portion of the filament just before the first eruption
showing the contraction of the loop over time. The white arrow shows the bright loop which was contracting. 
Top -- 2nd panel: Sequence of 171~\AA~images (a, b, c) (zoomed portion of the blue boxed region of 1st panel image) showing the 
contraction of the loop. 
Bottom -- 1st panel: Sequence of 171~\AA~ high-pass filtered images (A, B, C) for showing the contraction of the loop.
Bottom -- 2nd panel: Sequence of 171~\AA~high-pass filtered images (a, b, c) (zoomed portion 
of the blue boxed region of 1st panel image) showing the 
contraction of the loop.}
\label{fig:ch1_7}
\end{figure}

Just a few minutes before the eruption F1, there 
was an contraction of coronal loop on the eastern part of the filament. 
The loop contraction was observed during 2:32-2:35~UT. Figure~\ref{fig:ch1_7}~(top -- 1st panel)
shows the contraction of the coronal loop in a sequence of 171~\AA~images in different time epoch.
Magnified portion of the contracting loop regions (blue boxed region) are 
shown in Figure~\ref{fig:ch1_7}~(top -- 2nd panel) for the corresponding time of 
the top -- 1st panel images.
The time sequence of high-pass filtered 171~\AA~images for the contraction
 loop are shown in Figure~\ref{fig:ch1_7}~(bottom -- 1st panel) for better contrast.
 Magnified portion of the contracting loop regions (blue boxed region) 
 are shown in Figure~\ref{fig:ch1_7}~(bottom -- 2nd panel) for the corresponding time of 
the bottom -- 1st panel filtered images.
These high-pass filtered images are obtained by using the wavelet transform \citep{Young07}.
The contracted loop is shown by arrow in every images in both the panels.
During collapsing period it covered about 15$^{\prime\prime}$ in height. Later, though its signature is not 
visible due to the weakening of its brightness, we do see falling of material on to the filament 
at 2:37~UT indicating the collapsing loop reaching the filament location at that time. 
Soon after the collapse of the loop
the filament started to erupt.

\subsection{Simultaneous contraction and expansion of arcade loops}

\begin{figure}  
\begin{center}   
   \centerline{\hspace*{-0.01\textwidth}
               \includegraphics[width=1.10\textwidth,clip=]{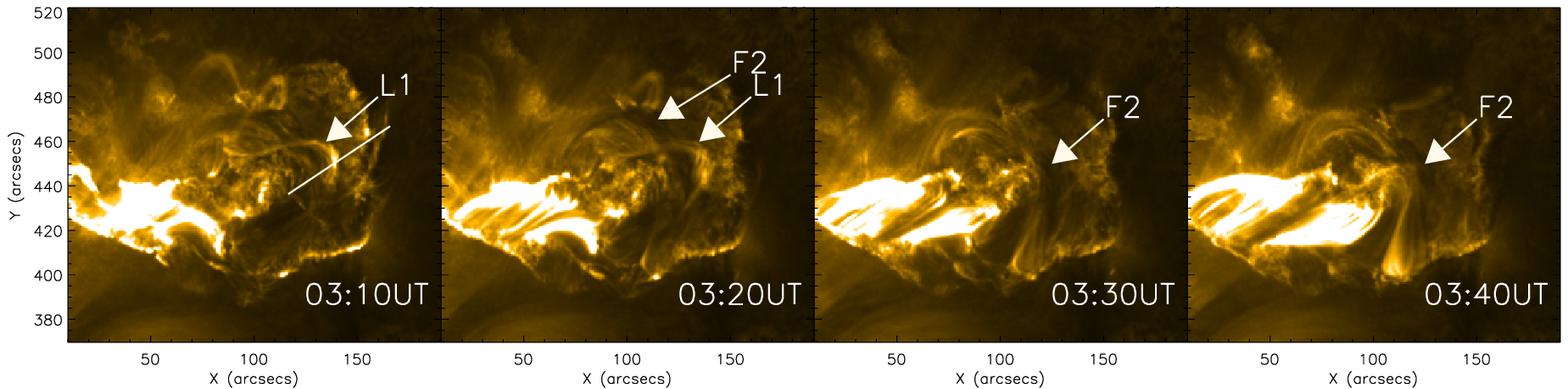}             
              }
              \vspace{0.02\textwidth}    
      \centerline{\hspace*{-0.03\textwidth}
               \includegraphics[width=1.10\textwidth,clip=]{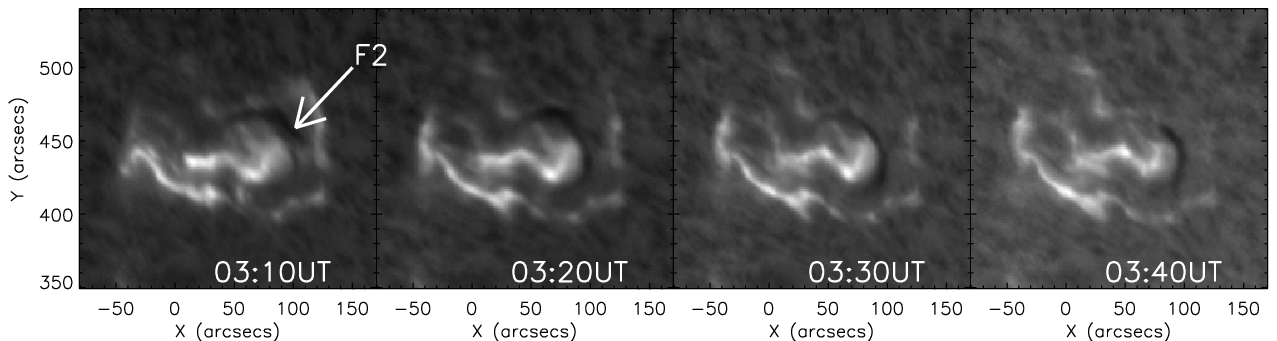}
               }
     \vspace{-0.05\textwidth}    
\end{center}        
\caption
{Top: Sequence of 171~\AA~images for the north-west portion of the filament after 
the first eruption showing the contraction
of one of the post-flare loop. The white arrow followed by `L1' shows the bright loop which was 
contracting before the second activation. The white arrow followed by `F2' indicate the filament location.
Bottom: Sequence of H$_{\alpha}$ images showing the filament position by white arrows (F2) with time.}
\label{fig:ch1_8}
\end{figure}

\begin{figure}    
 \begin{center}   
    \centerline{\hspace*{-0.01\textwidth}
                \includegraphics[width=0.55\textwidth,clip=]{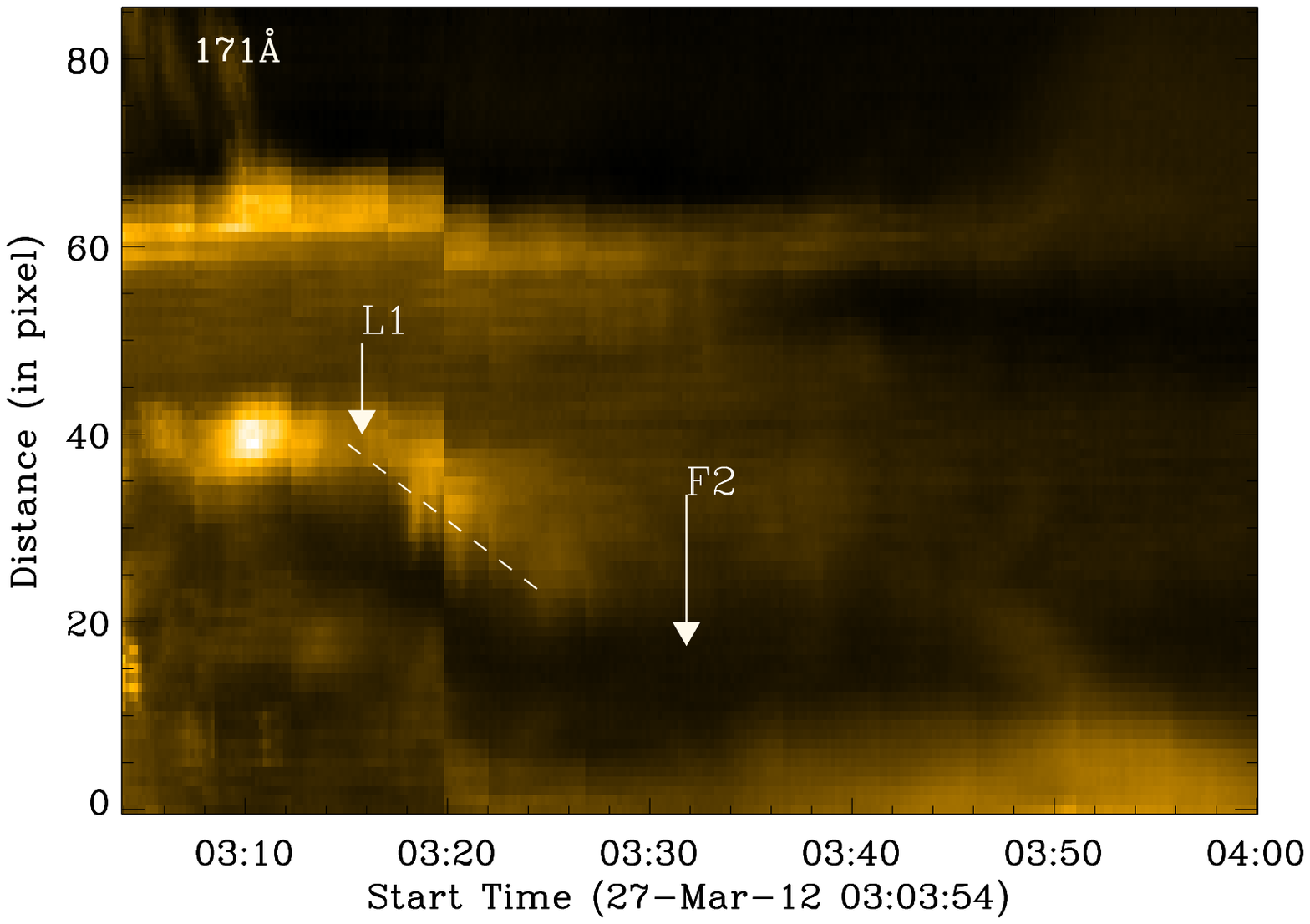}
                \hspace*{-0.07\textwidth}
               \includegraphics[width=0.55\textwidth,clip=]{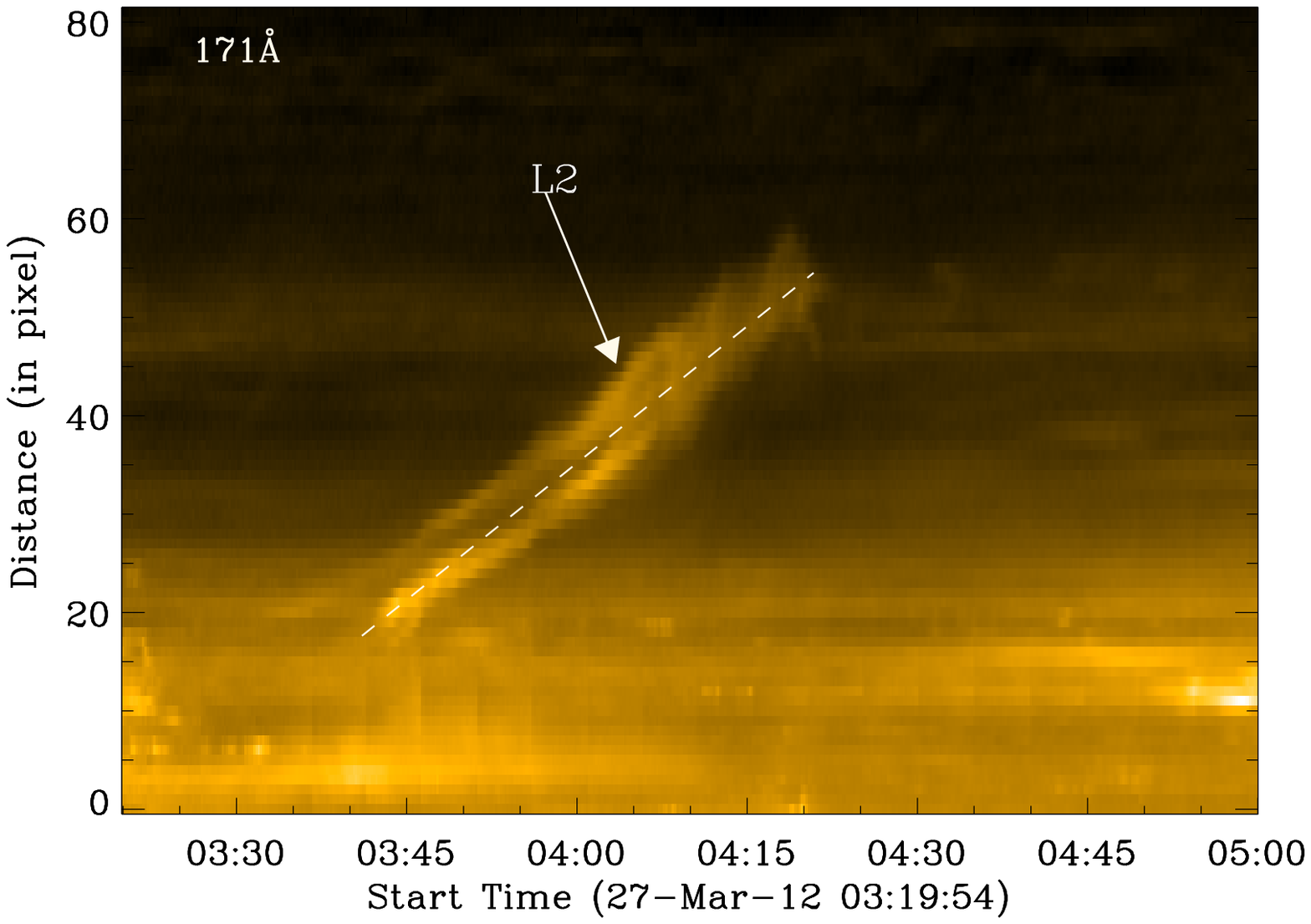}
            }           
      \vspace{-0.05\textwidth}    
 \end{center}        
 \caption{Left: The time-distance map of the contracting loop L1 along the slit of Fig.~\ref{fig:ch1_8} (top-1st image). 
 The filament and loop locations are shown
 using arrows followed by letter `F2' and `L1', respectively. The white dashed line 
 represents the path along which the loops are contracted.
 Right: The time-distance map of the expanding loop along the 
 slit of Fig.~\ref{fig:ch1_10} (Panel C). The white dashed line 
 represents the path along which the loops are expanded.}
 \label{fig:ch1_9}
 \end{figure}

After the first eruption (F1), the north-west portion of the filament activated. 
One of the post-flare loops started to shrink over the filament which is shown by an arrow (L1) in 
Figure~\ref{fig:ch1_8} (top). The space--time plot along the 
slit (overlaid on Figure~\ref{fig:ch1_8} (top-1st image)) shows the contraction
of the loop. The space--time plot is shown in Figure~\ref{fig:ch1_9} (left). The filament (F2) 
and contracting loop (L1) are shown in this plot. The contracting loop was also visible in the AIA 193, 211 and 335~\AA~channels.
The loop height reduced over the time. 
Once the shrinking loop
disappeared in the AIA 171~\AA~channel, the dark material in the filament appeared to be moving towards western side 
at 3:40~UT. After this, several loops around the filament cooled down and the 
length of the dark filament increased in size at 4:15~UT. 


%

In sequence of H$\alpha$ images, the shrinking loop was not visible, but the motion of the dark material
in the filament channel was visible over the time. At 3:20~UT, a dark arc shaped filament in the
northern portion of the two-ribbon flare was visible and is shown in Figure~\ref{fig:ch1_8} (bottom). 

\begin{figure}    
\begin{center}   
               \vspace{-0.02\textwidth} 
    \centerline{\hspace*{0.2\textwidth}
                \includegraphics[width=1.05\textwidth,clip=]{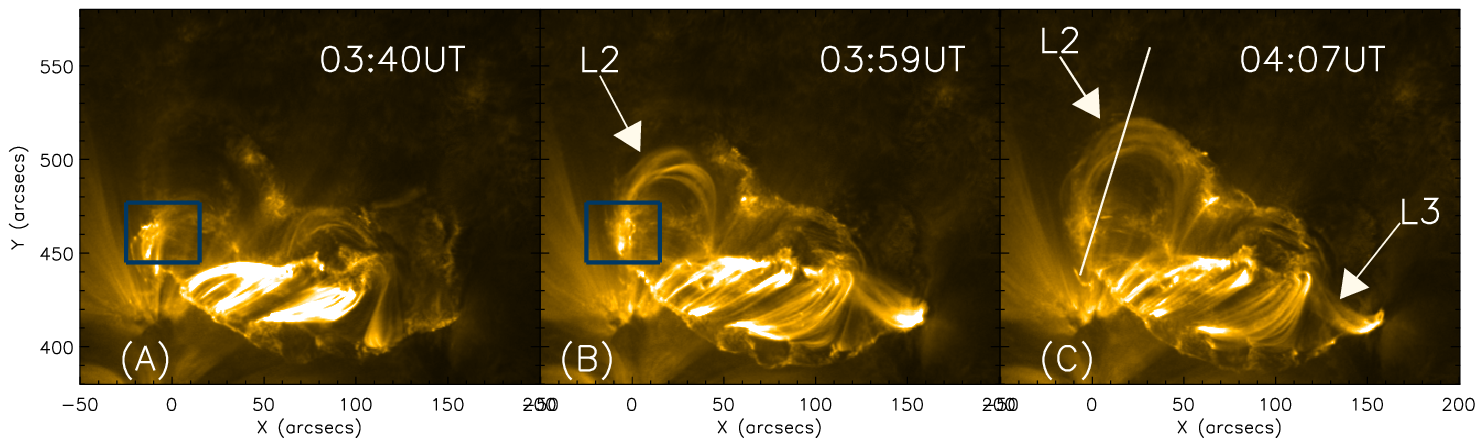}
               }
               \vspace{-0.05\textwidth}    
       \centerline{\hspace*{0.2\textwidth}
                \includegraphics[width=1.05\textwidth,clip=]{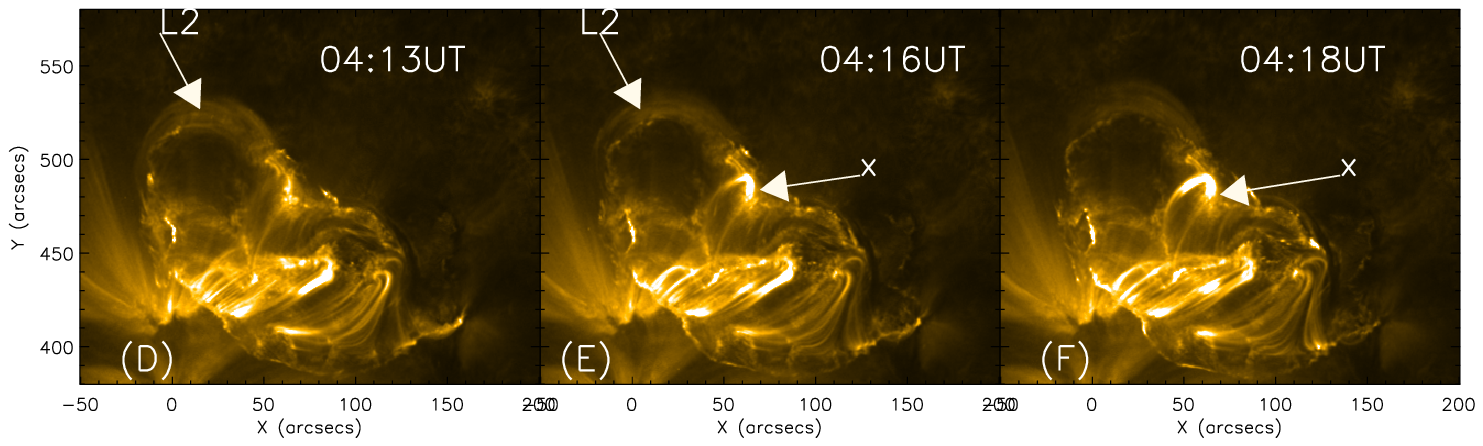}
                }
                   \vspace{-0.05\textwidth}    
       \centerline{\hspace*{0.2\textwidth}
                \includegraphics[width=1.05\textwidth,clip=]{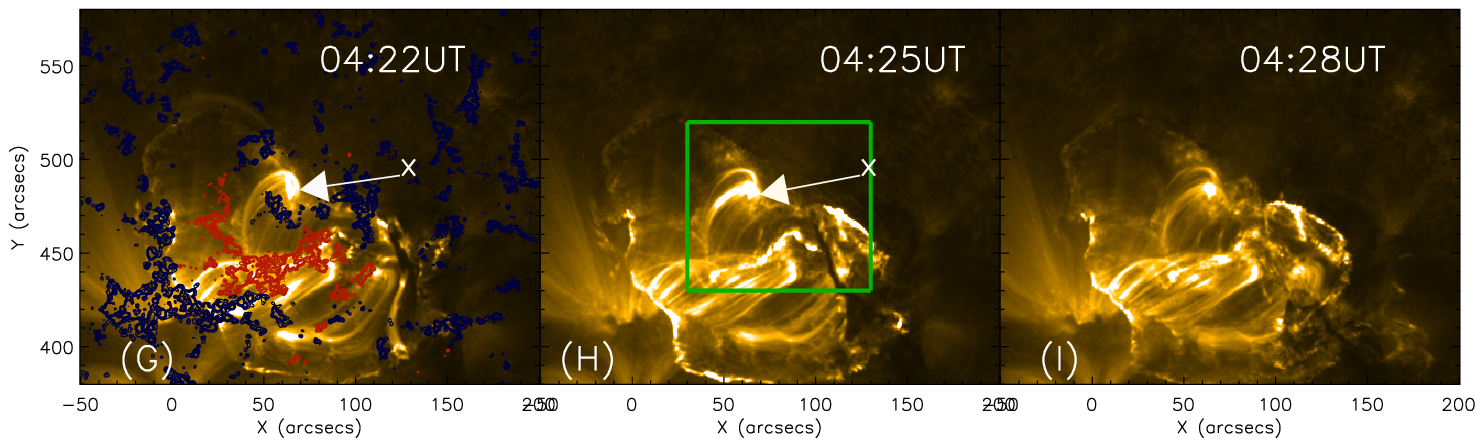}
                }   
                \vspace{0.01\textwidth}    
       \centerline{\hspace*{-0.01\textwidth}
                \includegraphics[width=0.8\textwidth,clip=]{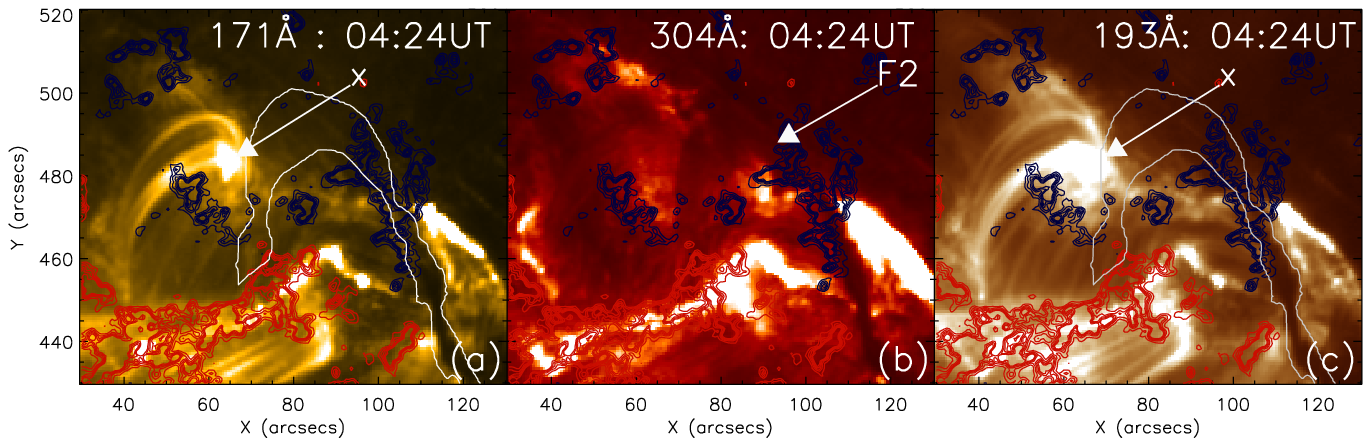}
                }
                \vspace{-0.055\textwidth}    
       \centerline{\hspace*{-0.01\textwidth}
                \includegraphics[width=0.8\textwidth,clip=]{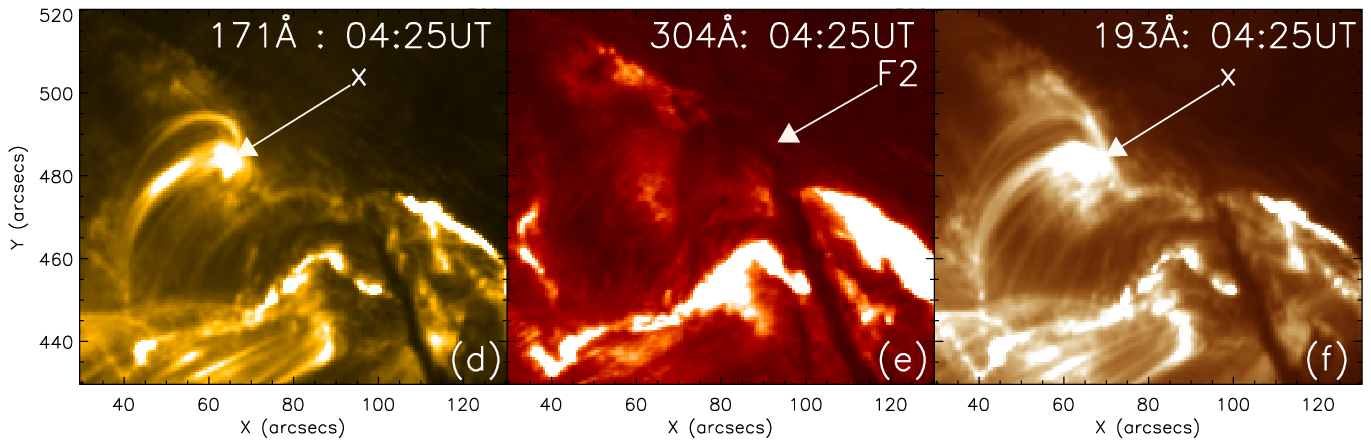}
                } 
      \vspace{-0.05\textwidth}    
 \end{center}        
 \caption{Panel (A)--(I): Sequence 
 of 171~\AA~images for the north-west portion of the filament before the 
second eruption showing the expansion and 
 collapse of coronal loops. The white arrow L2 and L3 show the expansion and 
 collapse of the coronal loops, respectively. 
 The white arrow `x' in the panels F--J shows the brightened loop close to eastern footpoint
 of the filament during filament activation.
 The blue boxed region (panel A \& B) shows the brightening close to eastern footpoint
 of the expanding loops. The red and blue contours on panel G represent the 
 positive and negative polarities with magnetic field strength values of $\pm$ 50, 100, 150, 200 
and 250~G, respectively.
 Panel (a)--(f): Magnified version of the green colored 
 boxed region of panel J image, are shown here in three
 different wavelengths 171, 304 and 193~\AA, respectively. The brightened 
 loops are shown by an arrow `x' in 171 and 193~\AA~ images 
  and the erupting filament is shown by arrow `F2' in 304~\AA~ image (panel b \& e). The erupting 
  filament contour overlaid on 171 and 193~\AA~images is extracted
 from 304~\AA~image.}
 \label{fig:ch1_10}
 \end{figure}

Near the east side of the filament, several loops were visible. At around 03:45~UT,
there was a radially outward expansion of these loops in the active region, as shown 
in a sequence of 171~\AA~images in Figure~\ref{fig:ch1_10}~(panel A--F). The arrow followed by `L2' shows the expanding loops.
A small brightening was observed at around
03:40 UT in EUV channel in the eastern footpoint location of the expanding loop. This
brightening location is shown by a blue rectangular box in Figure~\ref{fig:ch1_10}~(panel A \& B) in the 171~\AA~channel.

\begin{figure}  
\begin{center}   
                   \vspace{-0.02\textwidth}   
       \centerline{\hspace*{0.01\textwidth}
                \includegraphics[width=1.0\textwidth,clip=]{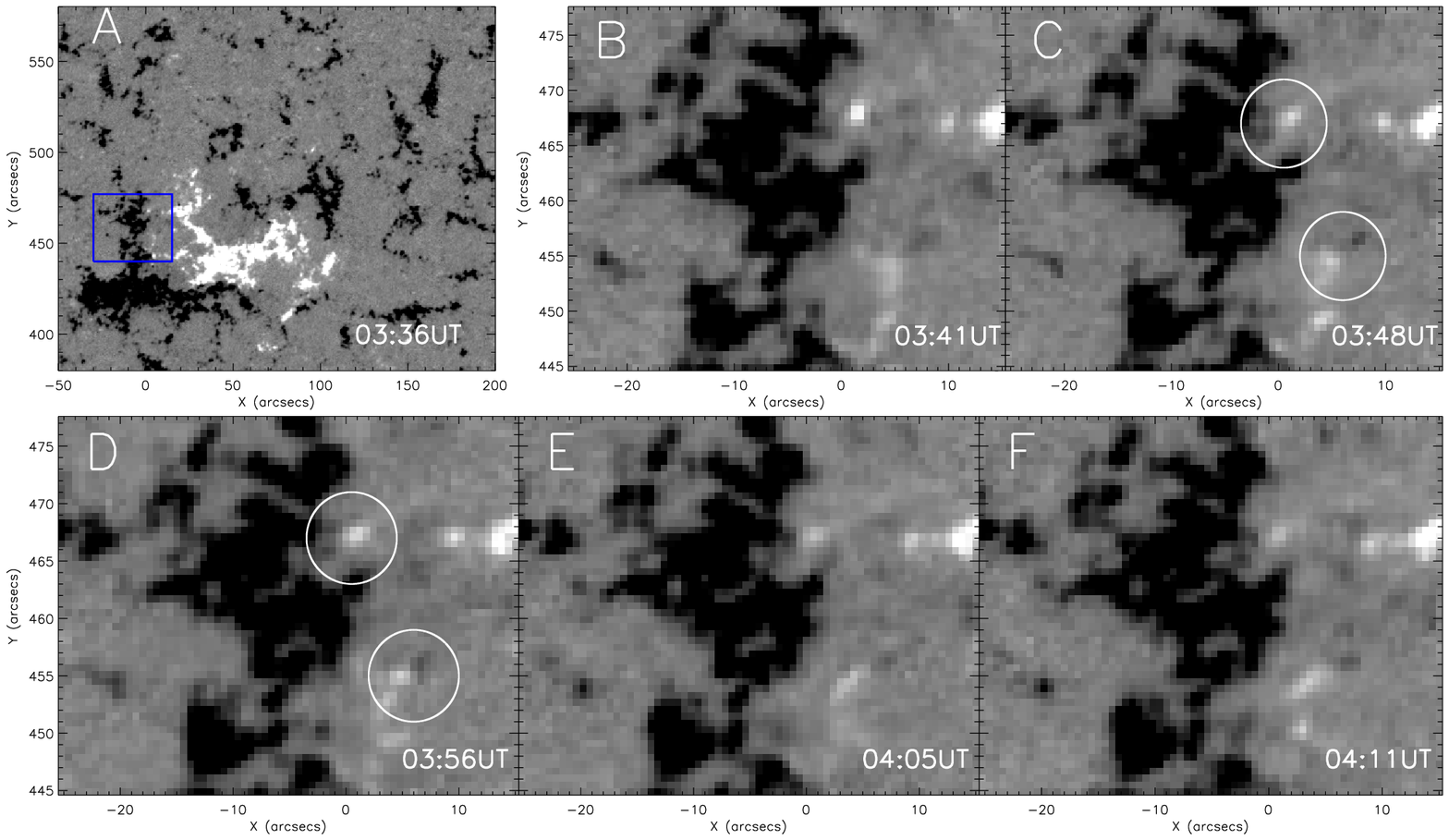}
                }
               \vspace{-0.07\textwidth}   
       \centerline{\hspace*{0.01\textwidth}
                \includegraphics[width=1.0\textwidth,clip=]{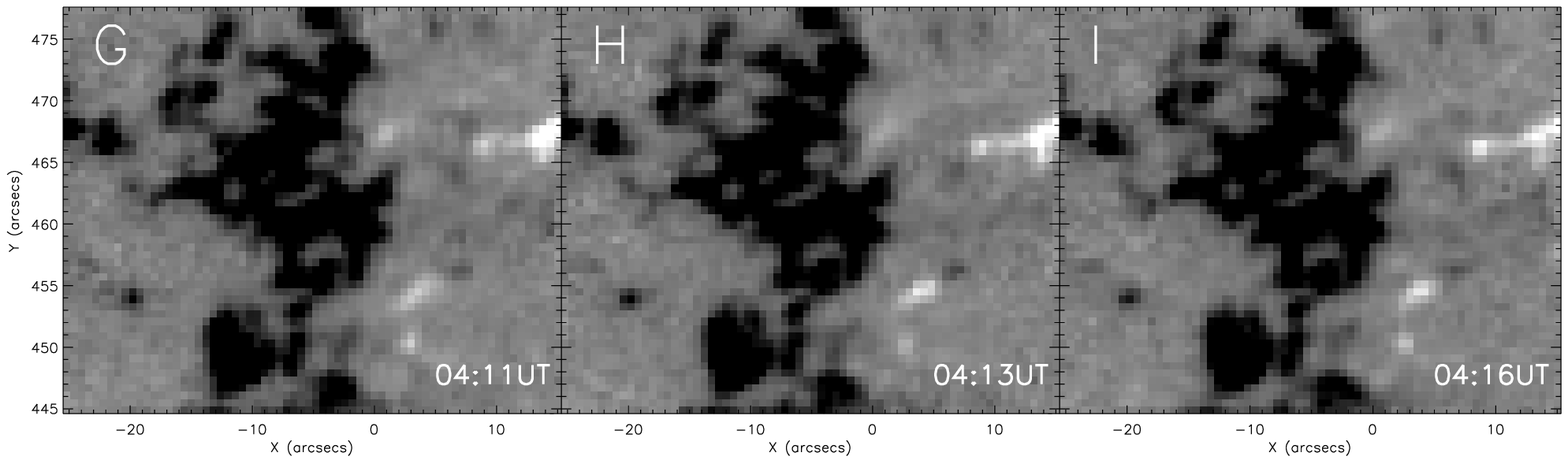}
                }             
      \vspace{-0.05\textwidth}   
 \end{center}        
 \caption{Panel A shows the line-of-sight magnetogram for the similar
 field of view of the top 171~\AA~ images of Figure~\ref{fig:ch1_10}. The boxed region is the same region as 
 shown in Figure~\ref{fig:ch1_10} (Panel A).
 The zoomed version of the boxed region of this magnetograms are displayed in a time 
 sequence in which circled regions indicate the region of flux cancellation (Panel B - I).
 Animation of the evolution of the magnetic field is available online (hmi.avi).
 }
 \label{fig:ch1_10a}
 \end{figure}

In the brightening location, a small-scale magnetic flux cancellation was observed
(shown in boxed region in Figure~\ref{fig:ch1_10a}~(panel A)). 
The magnified view of the box region in magnetograms is also shown in the
same Figure~\ref{fig:ch1_10a} (panel B -- I) with time. In the magnified magnetograms, the circled region show
the canceling features wherein the positive polarity region moved towards the negative
region and eventually canceled. 

During the time period 03:45~UT to 04:21~UT expanding loops reached about 60$^{\prime\prime}$ 
in height from initial position with a projected speed of 20$\pm$2.3~km/sec, 
while the collapsing coronal loops came down by 35$^{\prime\prime}$ with a projected speed of 12$\pm$2.0~km/sec.
 During the rise of the loop, a brightening was seen in the adjacent loop top. 
This brightened loop was cusp shaped, as shown by white an arrow `x'
in the sequence of 171~\AA~images in  Figure~\ref{fig:ch1_10} (panels F -- J).
The magnified  view of the green boxed region (shown in Figure~\ref{fig:ch1_10} (panel J))
are shown in 171, 304 and 193~\AA~wavelengths in the panels (a -- f) of Figure~\ref{fig:ch1_10}.
The brightened cusp shaped loop shown by arrow `x', is visible in 171 and 193~\AA~channels, but not in 304~\AA~channel.
The erupting filament is shown by arrow `F2' in 304~\AA~channel. The contour of the erupting filament 
extracted from 304~\AA~channel (panel b)
is overlaid on the 171 and 193~\AA~channels, shows the position of the filament.
This brightened loop was visible from $\sim$04:16~UT until $\sim$04:27~UT. The brightening could be due to 
the reconnection between the rising and overlying loops. 
At the same time the filament activated and started to rise. 
The top coronal loop on the western side of the filament started to collapse at the time $\sim$03:59~UT. 
The arrow followed by `L3' in Figure~\ref{fig:ch1_10}~(panel C \& D) shows the collapsing of the coronal loops.
Later, the filament rose to higher heights and the overlying loop in the western footpoint collapsed. 
While erupting, the filament appeared to be twisted and eventually disappeared.  
In summary, on the east side of the filament, the loops moved in 
upward direction and expanded. On the west side of the filament the loops moved in downward direction and shrunk.


\subsection{Magnetic field evolution}
\begin{figure}   
\begin{center}     
     \centerline{ \vspace*{0.00\textwidth}
               \includegraphics[width=0.66\textwidth,clip=]{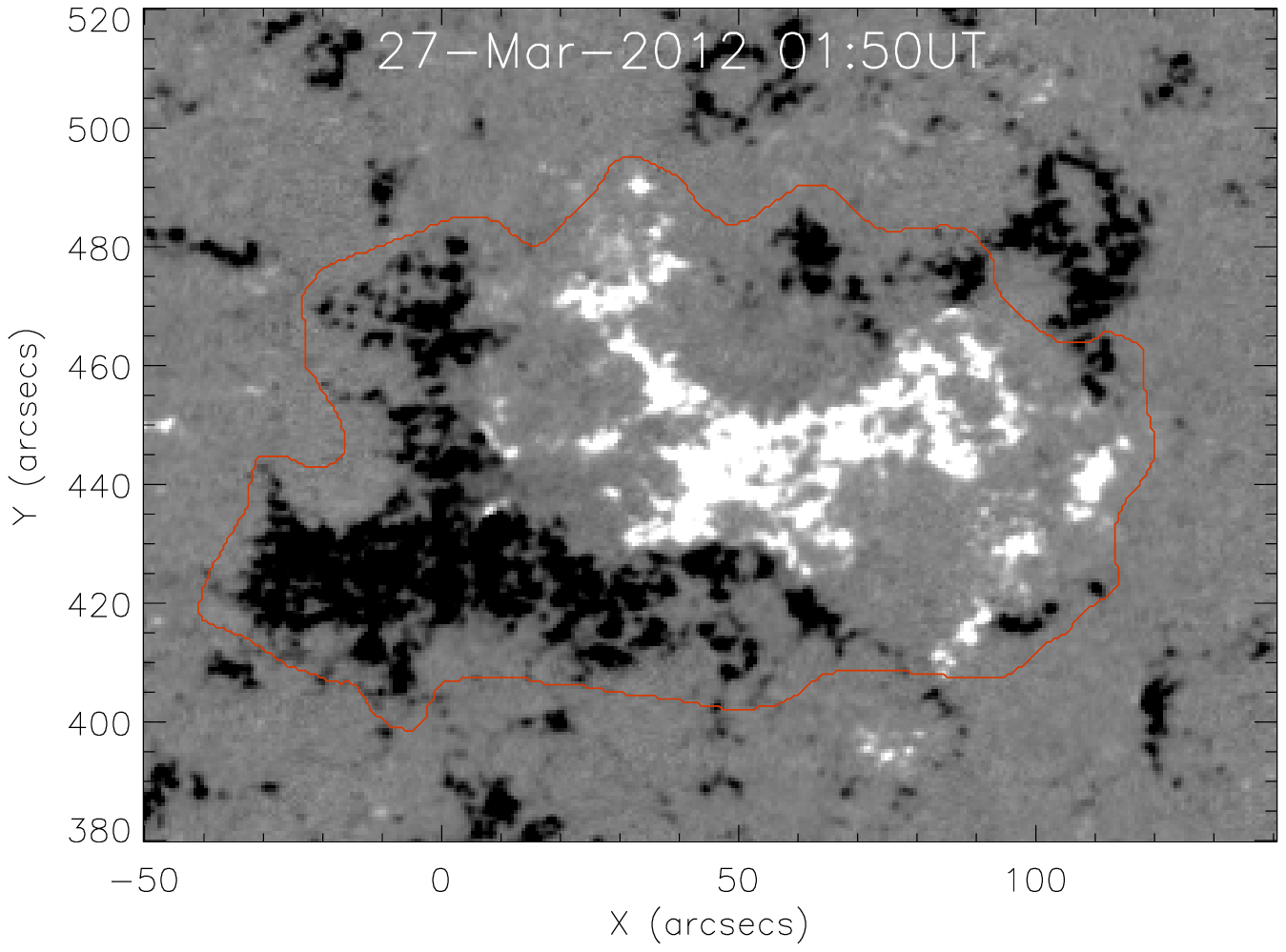}
               \hspace*{-0.17\textwidth}
               \includegraphics[width=0.66\textwidth,clip=]{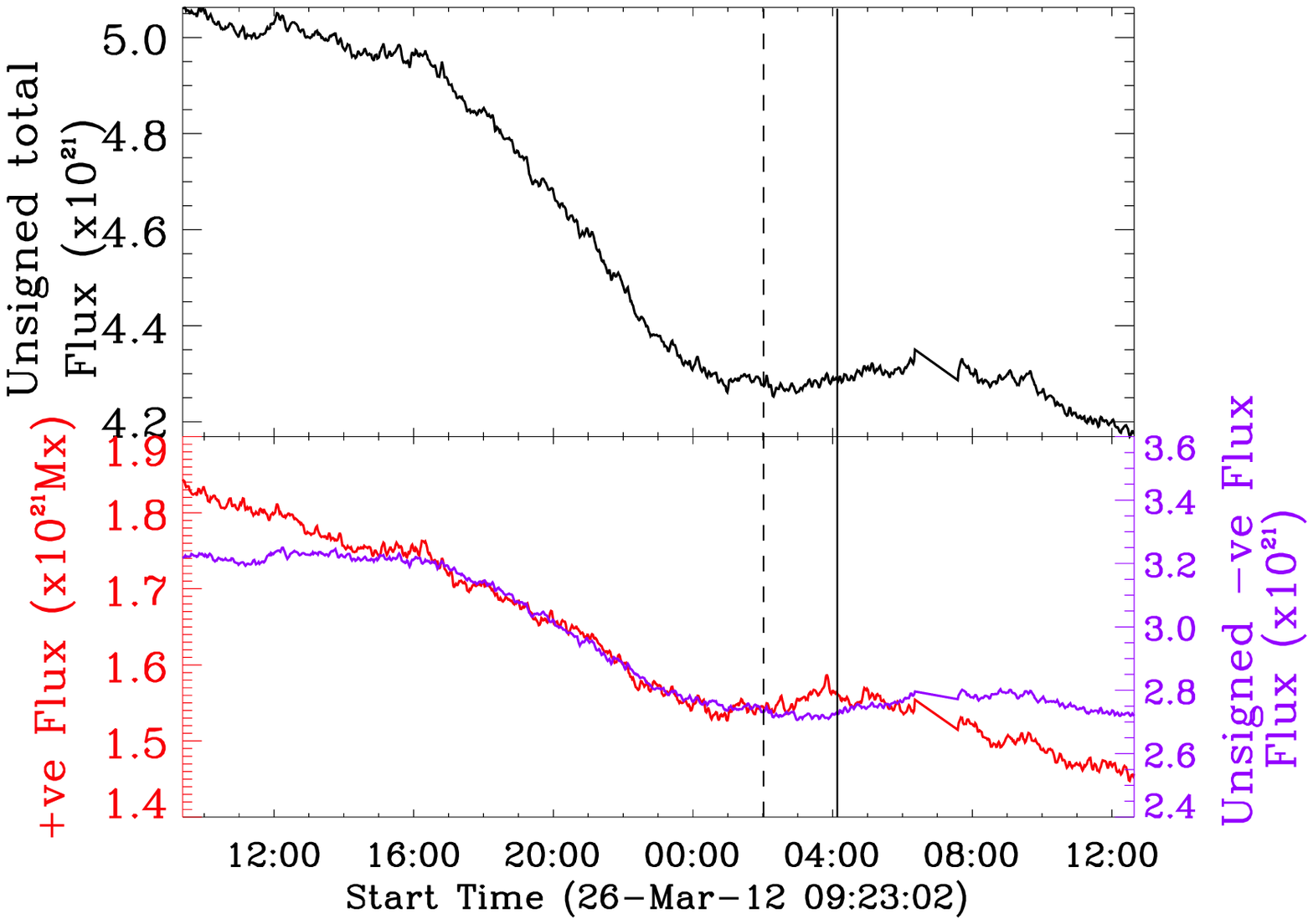}
              }
     \vspace{-0.05\textwidth} 
\end{center}        
\caption{
Left: Line-of-sight magnetogram of the region interest. Right: The plot 
shows the calculated positive (red), unsigned negative fluxes (blue), and unsigned total fluxes (black)
for the irregular contour. The dashed and solid
vertical lines represent the onset time of first and second filament eruptions, respectively.}

\label{fig:ch1_12}
\end{figure} 


The magnetogram shows the filament location near the PIL of the two opposite 
polarity plage regions (Figure~\ref{fig:ch1_1} (bottom-right)).  
The HMI magnetogram movie (hmi.avi) shows the interaction/cancellation of opposite polarity 
magnetic regions at the PIL.  
This region also showed a pre-flare brightening 
at around 01:56~UT on March 27, 2012.   
The flux cancellations near the PIL could be 
a cause for the preflare brightening near the filament.

Figure~\ref{fig:ch1_12} (left) shows the magnetogram few minutes before the pre-flare brightening. 
Magnetic flux in the active region was computed within the contoured region shown in Figure~\ref{fig:ch1_12}
(left). This contour encloses the bipolar region of interest. 
Figure~\ref{fig:ch1_12} (right) shows the calculated negative 
and positive fluxes for this contoured region. 
In the beginning there was a slight difference in the magnetic flux of both polarities.
The flux starts to decrease in both the polarities 
from the beginning of observations. However, in the negative polarity region the flux
started to decrease clearly after 16:00~UT. The flux decreased at a rate of $\sim$2.52$\times$10$^{19}$ Mx/hr  in the 
positive and  $\sim$5.89 $\times$10$^{19}$ Mx/hr in the negative polarity regions over 9 hours time period 
starting from 16:00~UT on March 26, 2012. When the flux decrease is stopped 
we observed the first filament eruption followed by the C5.3 class flare (shown by dashed line in the plot).  The 
second event occurred when there was a slight increase of negative flux and decrease 
in positive flux (shown by solid line in the plot).
The temporal evolution of the flux suggests that there was a continuous cancellations of the
flux in this region.

\begin{figure}   
\begin{center}    
     \centerline{ \vspace*{0.02\textwidth}
               \includegraphics[width=0.52\textwidth,clip=]{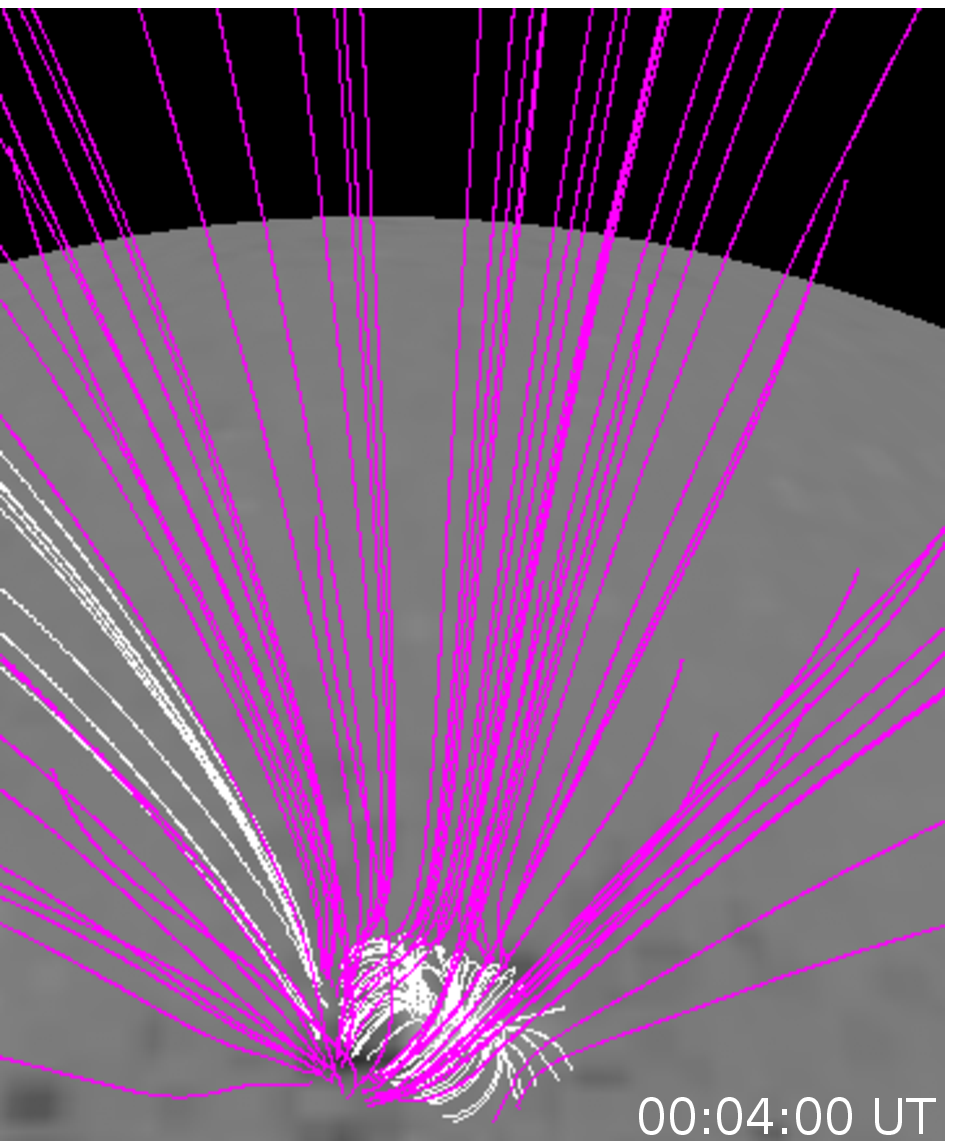}
               \hspace*{-0.005\textwidth}
               \includegraphics[width=0.46\textwidth,clip=]{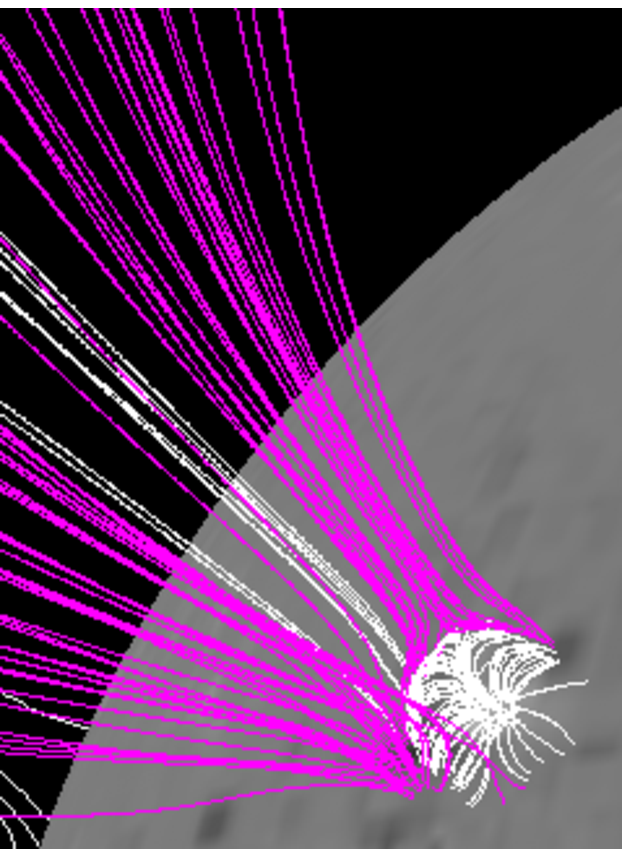}
               }
               \vspace{-0.01\textwidth}    
       \centerline{\hspace*{-0.001\textwidth}
                \includegraphics[width=0.49\textwidth,clip=]{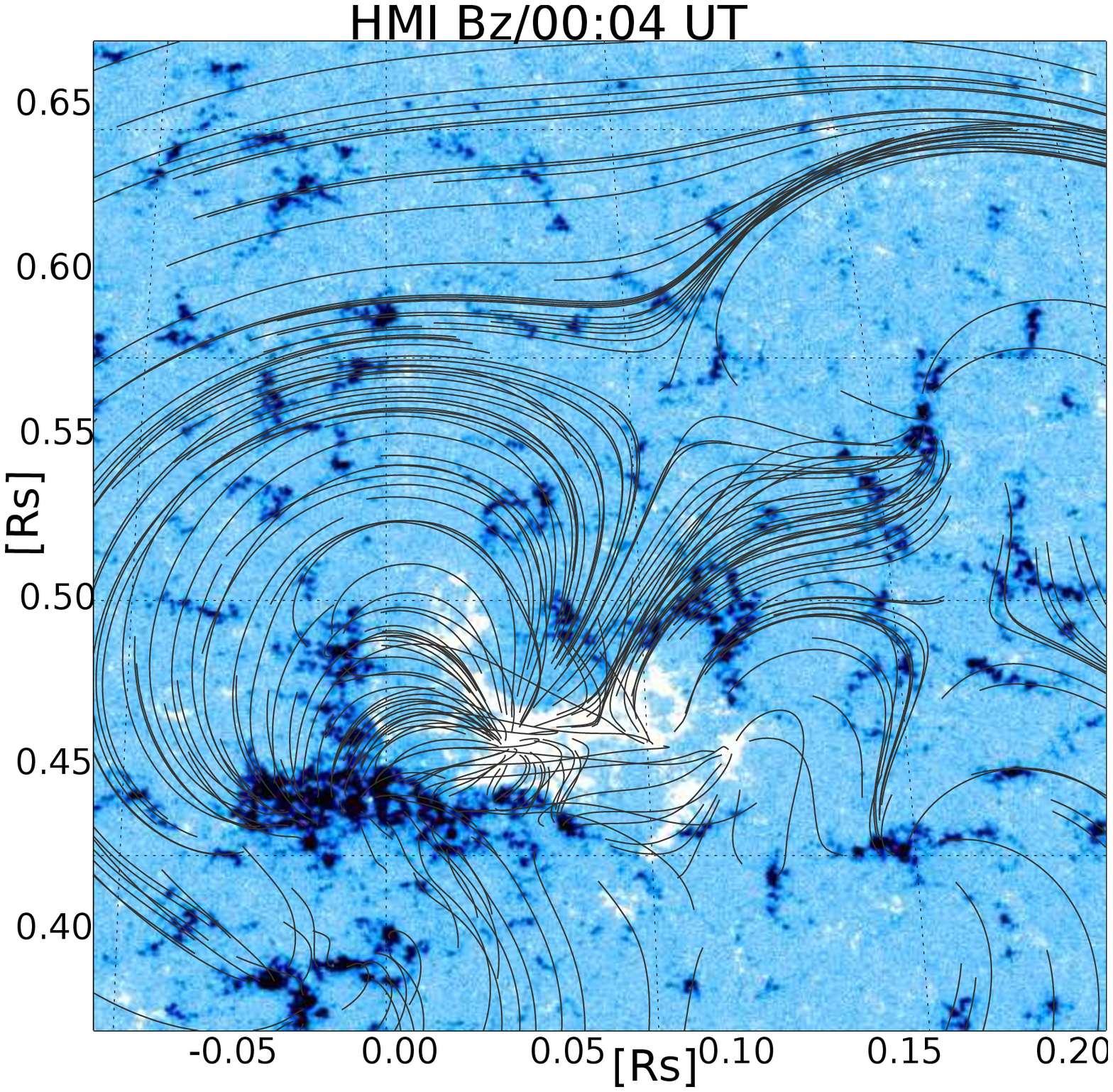}       
                \hspace*{-0.008\textwidth}
               \includegraphics[width=0.56\textwidth,clip=]{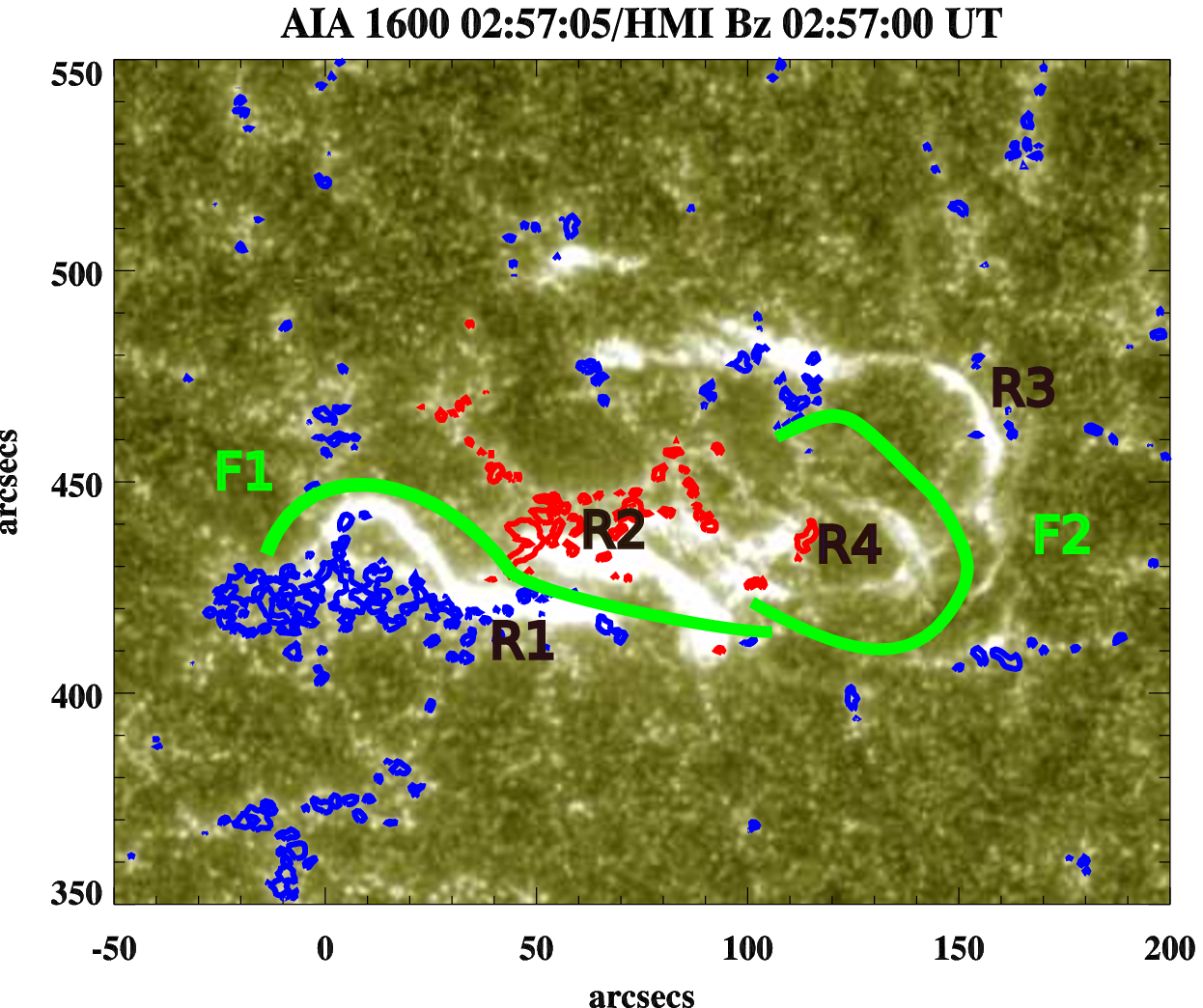}
              }
     \vspace{-0.05\textwidth} 
\end{center}        
\caption{Top-left: Potential field source surface extrapolation (PFSS) of the active region NOAA 11444 at 00:04:00 UT on
March 27, 2012.
The white and pink lines indicate the closed and open field lines. Top-right panel shows the rotated 
view of the AR (toward eastern limb). Bottom-left figure displays the NLFFF extrapolation of the field 
lines in a closer view. Bottom-right shows the AIA 1600 \AA~image during the appearance of flare 
ribbons by the eruption of F1 at 02:57 UT. The flare ribbons are shown by R1, R2, R3 and R4.}
\label{fig:pfss1}
\end{figure} 

To explore the magnetic topology of the active region, we performed the potential 
field extrapolation using a HMI magnetogram at 00:04:00 UT on March 27, 2012. Figure~\ref{fig:pfss1} (left panel) 
displays the potential field source surface extrapolation (PFSS; \opencite{Schrijver03}) of the active region. 
The white and pink lines indicate the closed and open field lines. The rotated view of the AR (toward eastern limb) 
is shown in the right panel. The fan loops emanates from the central positive polarity region and connects the 
surrounding opposite polarity fields.
The magnetic configuration is almost similar to a fan-spine topology \citep{Pariat2010} with a possible magnetic null point.

We used ``Vertical-Current Approximation Nonlinear Force-Free Field" (VCA - NLFFF; \opencite{Aschwanden2016}) code using 
SDO/HMI and SDO/AIA  data to obtain a nonlinear force-free field solution of the magnetic field in 
an active region. 
We used HMI line of sight magnetogram obtained at 00:04 UT and near simultaneous 6 channels AIA images (211, 193, 94, 
131, 335 and 171 \AA) as an input to determine the NLFFF solution. The bottom-left panel 
of Fig.~\ref{fig:pfss1} displays the NLFFF extrapolation of the field lines in a closer 
view. We can see clear connectivities of the central positive polarity region to the 
surrounding negative polarities, which is in agreement with PFSS extrapolation.
It seems to have a magnetic null point above the positive polarity region, which is expected in a breakout magnetic topology.

The right panel shows the AIA 1600 \AA~image during the appearance of flare 
ribbons by the eruption of F1 at 02:57 UT. We see multiple flare ribbons (R1, R2, R3,and R4). We draw the 
location of the filaments F1 and F2 to see their association with flare ribbons. Ribbons R1/R2 are associated 
with F1 whereas R3/R4 with F2. However, F2 could not erupt during the first flare and stopped by the overlying 
loop systems as shown in NLFFF extrapolation.

We notice low-lying connecting loops above the filament F1 and no higher 
loops are revealed at the sight of F1. However, we see higher closed fields above F2 where 
it was trapped during the first flare. Therefore, NLFFF extrapolation suggests the role of overlying 
magnetic fields in producing the successful and failed eruption.


\begin{table}[h!t]

 \tiny

\vspace{0.2in}
\caption{Summary of the events.}

\begin{tabular}{l|p{8cm}}

 \hline

Time (UT) & Observation  \\

 \hline

$\sim$01:56                &   A brightening was observed (green boxed region in Figure~\ref{fig:ch1_1} (top-right))
                              in AIA 171, 193, 131 and 304~\AA channels. \\ 
$\sim$01:57 -- $\sim$02:35 &   A bright flow was observed, which moved from east side of the filament 
                              to the west side. \\ 
$\sim$02:30                &   One more brightening was observed near the east end of the filament which is 
                              related to preflare brightening of C5.3 class flare (see Figure~\ref{fig:ch1_5}) and filament 
                              eruption (F1) starts.\\ 
$\sim$02:32 --$\sim$02:35  &   Contraction of coronal loop on the eastern part of the filament (see Figure~\ref{fig:ch1_7}).\\                        
$\sim$02:53                &   C5.3 class flare along with the filament eruption (F1).  \\ 
$\sim$03:08                &   Peak time of the C5.3 class flare. \\ 

$\sim$03:14                &   A dark filament portion appeared over the bright and dark filament channel 
                              (see Figure~\ref{fig:ch1_8}). \\ 
$\sim$03:14 --$\sim$03:40  &   The north-west portion of the filament started to activate. Shrinking of one of the
                              post-flare loop (L1) is visible (see Figure~\ref{fig:ch1_8}). \\ 
                           
$\sim$03:45                &   There was an expansion of loops (L2) towards higher heights in the east side of
                              filament. At the same time a contraction of loops (L3) in the west side are also
                              visible and these loops were sitting over the filament (see Figure~\ref{fig:ch1_10}).\\  
$\sim$04:16 -- $\sim$04:26 &   A brightened cusp shaped loop was observed near the east end of the filament, 
                              as shown by arrow `x' in the Figure~\ref{fig:ch1_10}~(panel a -- f).\\                        
                           
$\sim$04:20                &   This filament eruption (F2) started followed by C1.7 class flare.\\

 \hline 

\end{tabular}
\label{Table:1}

\end{table}


\section{Summary and Discussions} 
      \label{section4} 

We observed successive eruption of filaments (F1 \& F2) in AR NOAA 11444. The inverse `J' shaped filament
erupted in two different phases. In the first phase of the eruption the southern portion of the filament erupted,
while in the second phase the northern portion erupted.
The summary of the events are given in the Table~\ref{Table:1}. These two phases of eruptions were 
accompanied with GOES C5.3 and C1.7 flares, respectively. 
     
The transient brightening in EUV started at $\sim$01:56 UT at bottom of the filament, as discussed in Section~3.2.
and it was followed by a bright plasma flow which moved from
east end of the filament to the west. The filament appeared as a
sequence of bright and dark threads in 171~\AA~during this flow.
Photospheric magnetograms 
showed continuous cancellation of the flux near the PIL. 
The flux cancellation produced a jet below the filament F1. 
\cite{Chae03b} studied the formation of reverse S-shaped 
filament in NOAA AR 8668 associated with series of magnetic flux cancellation at the photosphere.
The flux cancellations 
along the PIL can form the twisted helical flux rope which support the filament, as formulated by \cite{VanBallegooijen89}
and
the same processes can also destabilize the filament \citep{Amari03,Martin92}.
Prior to eruption, the filament F1 rose slowly at a velocity of 1.5$\pm$0.3 km~s$^{-1}$. 
The filament's slow rise started about concurrently with the EUV brightening beneath the F1. 
Later, it accelerated during the flare.  
A similar kind of slow rise of filament prior to eruption during the EUV brightening beneath the 
filament was reported by \cite{Sterling05}. They interpreted their observations as  
the slow-rise phase of the eruption is resulted from the onset of tether cutting reconnection beneath the filament. 
We believe that the slow rise of the filament in the first phase of the eruption was most likely caused by 
the tether cutting reconnection \citep{Moore92,Sterling05} due to the flux cancellation 
between the opposite polarity magnetic fields below F1.
The reconnection below F1 can reduce the tension of the 
overlying twisted fields and push F1 to higher heights.

\inlinecite{Sterling11} observed a gradual magnetic flux cancellation under the filament, which built the 
filament flux rope over the time, causing it to rise gradually. The filament eventually erupted due to 
the onset of a magnetic instability and/or runaway tether cutting.
During the onset and early development of the explosion for six bipolar events, \inlinecite{Moore01} observed that 
in each of the events the magnetic explosion was unleashed by runaway tether-cutting via 
reconnection in the middle of the sigmoid. As this internal reconnection starts at the beginning  
of the sigmoid eruption and grows in step of the magnetic explosion, 
their study suggests that this reconnection is essential for the onset and growth of 
magnetic explosion in eruptive flares.
Recent results also show a direct observational evidences for tether-cutting mechanism 
in at least two events \citep{Chen14}. We believe that the observed brightening near the low lying 
filament end followed by a jet-like flow along the filament could be the
reconnection occurred in the lower layers and a signature of tether-cutting mechanism, which could  
have activated the filament F1 and it eventually erupted during C5.3 class flare.

Further, in the same active region, we observed a second phase of filament eruption F2, 
after $\sim$1.5~hrs of the first filament eruption (F1). 
We observed a flux cancellations near the east side of the filament wherein a 
small positive polarity region interacts with the negative polarity region.  
A brightening, followed by rising of the loops to higher heights was observed at the same 
location. 
The eruption of the overlying loops above
filament could have removed the sufficient 
amount of overlying flux (as numerically explained by \inlinecite{Torok11}), and made an easy way for the filament eruption F2.
For the same event, \cite{Lee2016} also explained that the eruption was possible as the topology of the dome-like magnetic
fan structure confined over the AR, was enabled to continue during the eruptions in spite of the significant changes in its geometry.
The reason for the collapsing loop observed near the western end of the filament is not clear. However, \inlinecite{Liu10,Yan13}
explained that the collapsing of the loops could be
due to the reduced magnetic pressure underneath the filament.

In summary, the filament eruption F1 is
consistent with the scenario: the gradual flux cancellation at PIL resulted the brightening 
followed by jet-like ejection below filament main
axis. The brightening could be due to the reconnection
occurred in the lower layers and a signature of tether-cutting mechanism, which could have triggered filament F1 initially with slow-rise
and it eventually erupted during C5.3 class flare.
However, filament F2 was trapped within the overlying loops and most likely did not 
reach the height of torus instability/loss of equilibrium and remain trapped for about 1.5 hrs. The 
filament F2 erupted after the removal of the overlying loops.
Both filament eruptions were associated with two high 
speed CMEs (\footnotemark[1]1148~km/s and \footnotemark[2]880~km/s, respectively),
which were observed by SOHO/LASCO and STEREO COR1 \& COR2 coronagraphs.
The future studies  of successive filament eruption with high resolution observations would be helpful to understand 
the origin of sequence of events that trigger the eruption.

\footnotetext[1]{http://cdaw.gsfc.nasa.gov/CME$\_$list/UNIVERSAL/2012$\_$03/htpng/20120327.031209.p346g.htp.html}
\footnotetext[2]{http://cdaw.gsfc.nasa.gov/CME$\_$list/UNIVERSAL/2012$\_$03/htpng/20120327.042405.p097g.htp.html}

\begin{acks}
We thank the referee for many valuable and
insightful comments which greatly helped us to improve the quality of the manuscript.
The AIA data used here is the courtesy of SDO (NASA) and AIA consortium. SDO/HMI 
is a joint effort of many teams and individuals to whom we are greatly indebted for 
providing the data. This work also utilizes data obtained by the Global Oscillation Network Group (GONG) Program, 
managed by the National Solar Observatory, which is operated by AURA, Inc. under a cooperative 
agreement with the National Science Foundation. The data were acquired by instruments operated by 
the Big Bear Solar Observatory, High Altitude Observatory, Learmonth Solar Observatory, Udaipur Solar Observatory, 
Instituto de Astrofisica de Canarias, and Cerro Tololo Inter-American Observatory. The EUVI images
are supplied courtesy of the STEREO Sun Earth Connection Coronal and Heliospheric Investigation (SECCHI) team.
\end{acks}

\end{article} 

\end{document}